\documentclass[3p]{elsarticle}

\usepackage{bigints}
\usepackage{natbib}
\usepackage{color}
\usepackage{url}
\graphicspath{{./figures/},{./figures_mcr_quad/},{./figs_legends/}}
\DeclareGraphicsExtensions{.pdf,.png,.jpg,.eps}
\usepackage{hyperref}

\usepackage[version=3]{mhchem} 

\begin{document}

\begin{frontmatter}

\title{Equilibrium microstructures of diblock copolymers under 3D confinement} 
\author[1]{Ananth Tenneti}
\author[1]{David M Ackerman}
\author[1]{Baskar Ganapathysubramanian\corref{cor1}}
\ead{baskarg@iastate.edu}
\address[1]{Department of Mechanical Engineering, Iowa State University, Ames, IA, United States}
\cortext[cor1]{Corresponding author}

\begin{abstract}

We investigate equilibrium microstructures exhibited by diblock copolymers in confined 3D geometries. We perform Self-Consistent Field Theory (SCFT) simulations using a finite-element based computational framework (Ackerman et al.\cite{ACKERMAN2017280}), that provides the flexibility to compute equilibrium structures under arbitrary geometries. We consider a sequence of 3D geometries (tetrahedron to sphere) that have the same volume but exhibit varying curvature. This allows us to study the interplay between edge and curvature effects of the 3D geometries on the equilibrium microstructures. We observe that beyond a length scale, the equilibrium structure changes from an interconnected network to a multi-layered concentric shell as the curvature of the 3D geometry is reduced. However, below this length scale the equilibrium structure remains a multi-layered concentric shell independent of curvature. We additionally explore variations in the equilibrium microstructures at a few discrete volume fractions. This study provides insight into possible frustrated phases that can arise in AB diblock systems while varying the shape of confinement geometry.
\end{abstract}

\end{frontmatter}

\section{Introduction}
Self-assembly of block copolymer systems leads to diverse equilibrium structures with a wide range of applications in  optics~\citep{yabu2011},  electronics~\citep{B815166K}, photonics~\citep{doi:10.1002/adma.200290018} and drug delivery systems~\citep{CHAN20091627}. Motivated by these applications, the equilibrium microphase structures of multi block copolymers have been investigated via experiments and simulations \citep{doi:10.1021/ma101360f,doi:10.1021/ma00186a051,PhysRevE.65.030802,doi:10.1021/ma202542u,C1SM05747B,doi:10.1021/nn101121n,Thompson2469,doi:10.1021/jacs.5b00493}. The equilibrium structures depend on polymer composition, interaction strength, system size, and shape of the confining geometry. The microphases and phase transitions of bulk diblock systems have been previously characterized \citep{{doi:10.1021/ma951138i},{doi:10.1021/ma00130a012},{BATES898}}. However, there has been an increasing interest in equilibrium microstructures formed by polymers under confinement. Confinement induces a wide variety of microphase structures which cannot be realized in bulk \cite{C3SM52821A}. Using numerical simulations, diblock systems under spherical, cylindrical and polyhedral confinement have been studied \citep{{doi:10.1021/ma071624t},{doi:10.1063/1.3050102},{doi:10.1063/1.2735626},{C1SM05947E},{doi:10.1021/la200379h}}. These results suggest the possibility that edge effects in polyhedral shapes can lead to stronger confinement effects as compared to spherical shapes \citep{C3SM52821A}. This is the primary motivation for the current work. More broadly, a deeper understanding of the interplay between confinement, edge effects and curvature on the equilibrium microstructures is valuable for tailoring the equilibrium morphology. This is especially promising due to recent advances in manufacturing that allow generating micro and nano particles with arbitrary 3D geometries. Promising techniques include chemical self-assembly \citep{doi:10.1021/nl052409y}, DNA templating \citep{Han342}, electron beam lithography \citep{doi:10.1021/nl0505492} and more recently, inertial microfluidics for flow sculpting and arbitrary shape design \citep{Amini2013}. 

\begin{figure*}
\begin{center}
\includegraphics[width=1.2in]{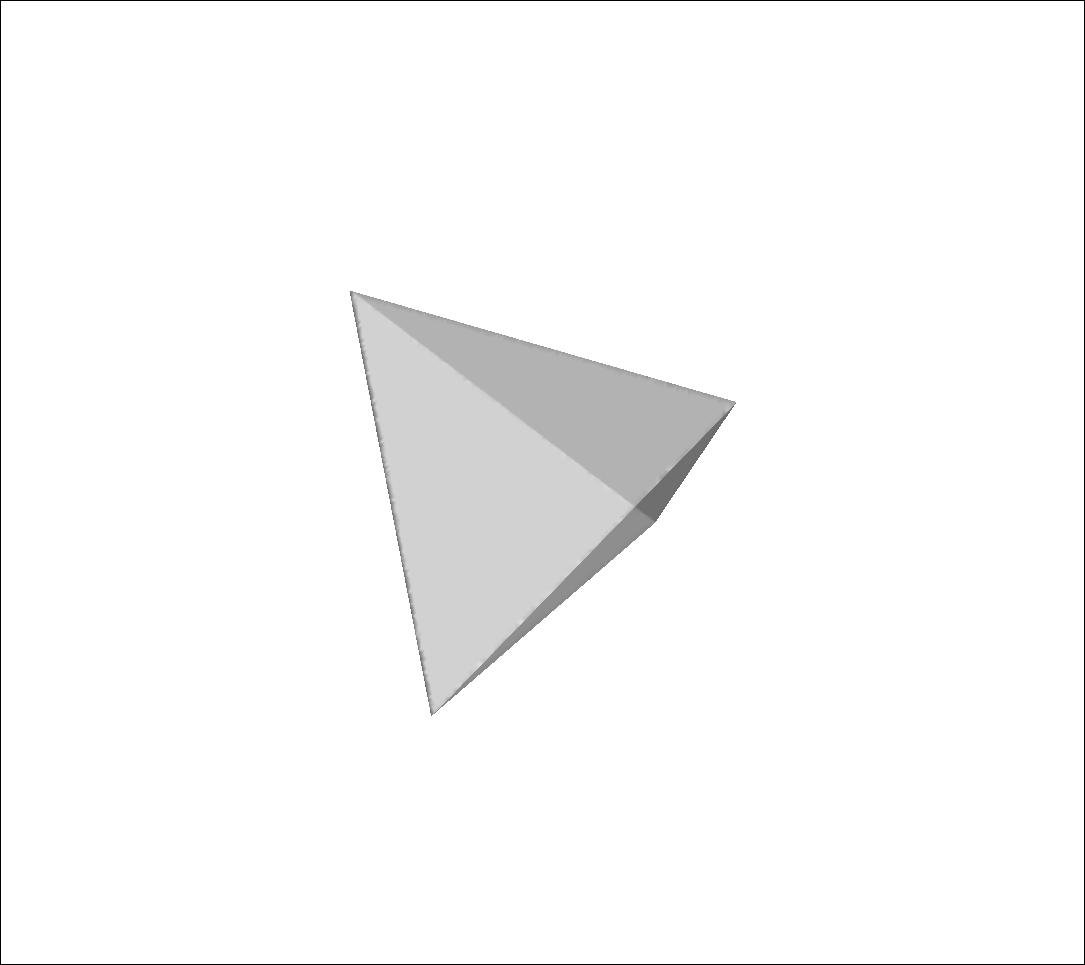} \hspace*{-0.9em}
\includegraphics[width=1.2in]{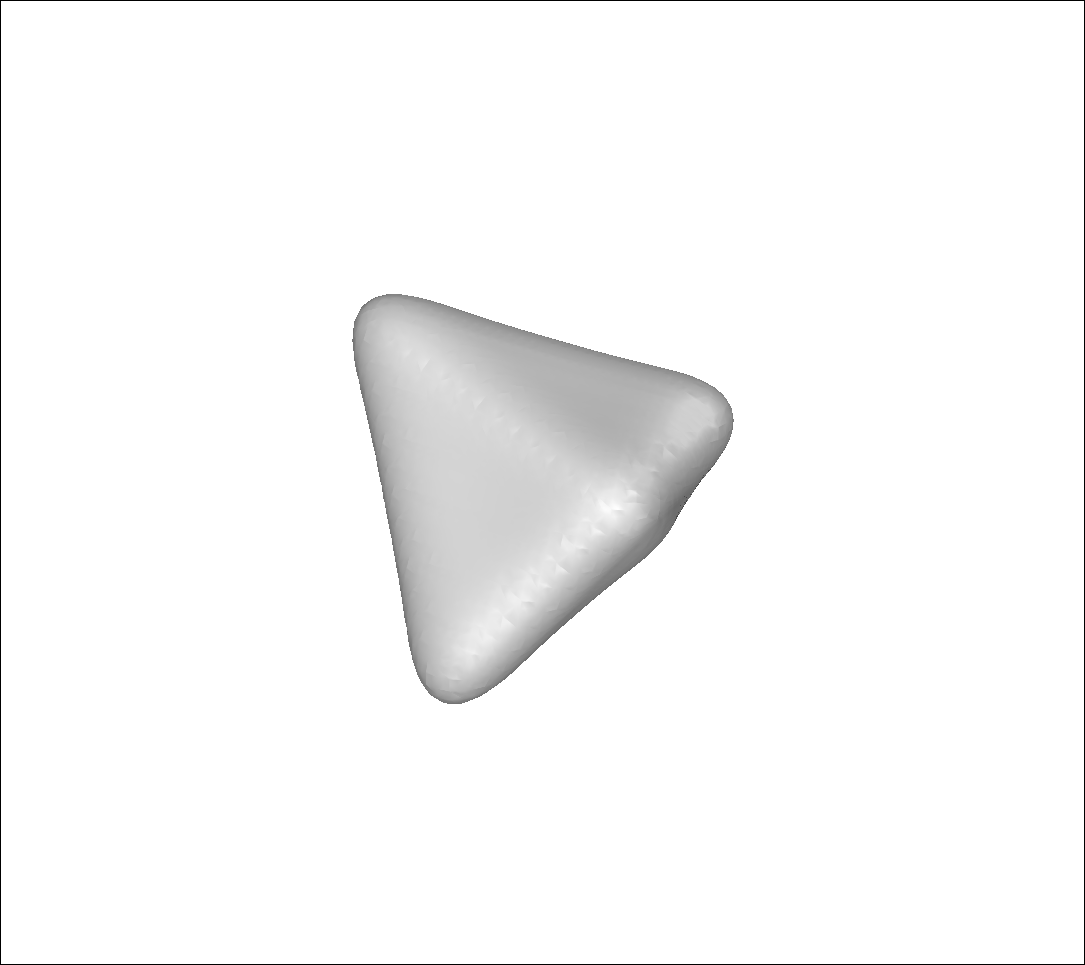} \hspace*{-0.9em}
\includegraphics[width=1.2in]{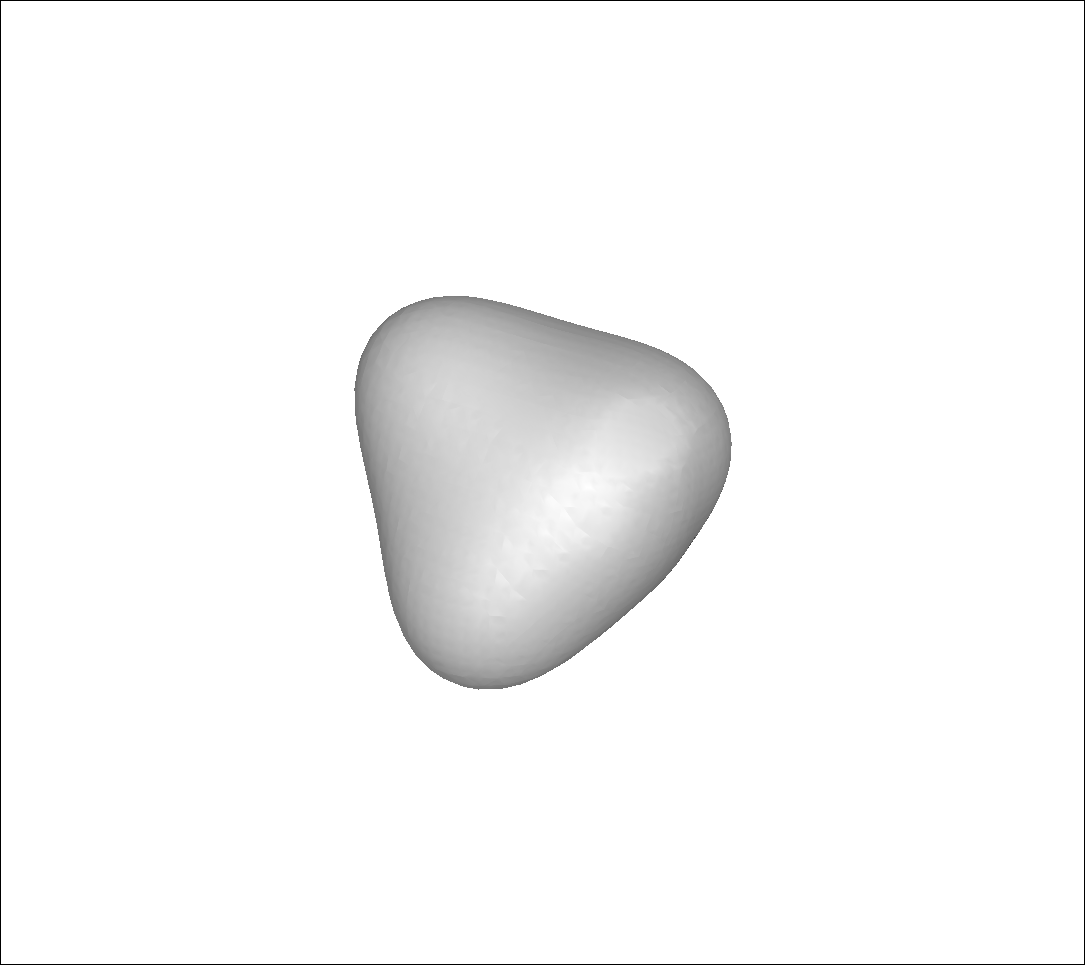} \hspace*{-0.9em}
\includegraphics[width=1.2in]{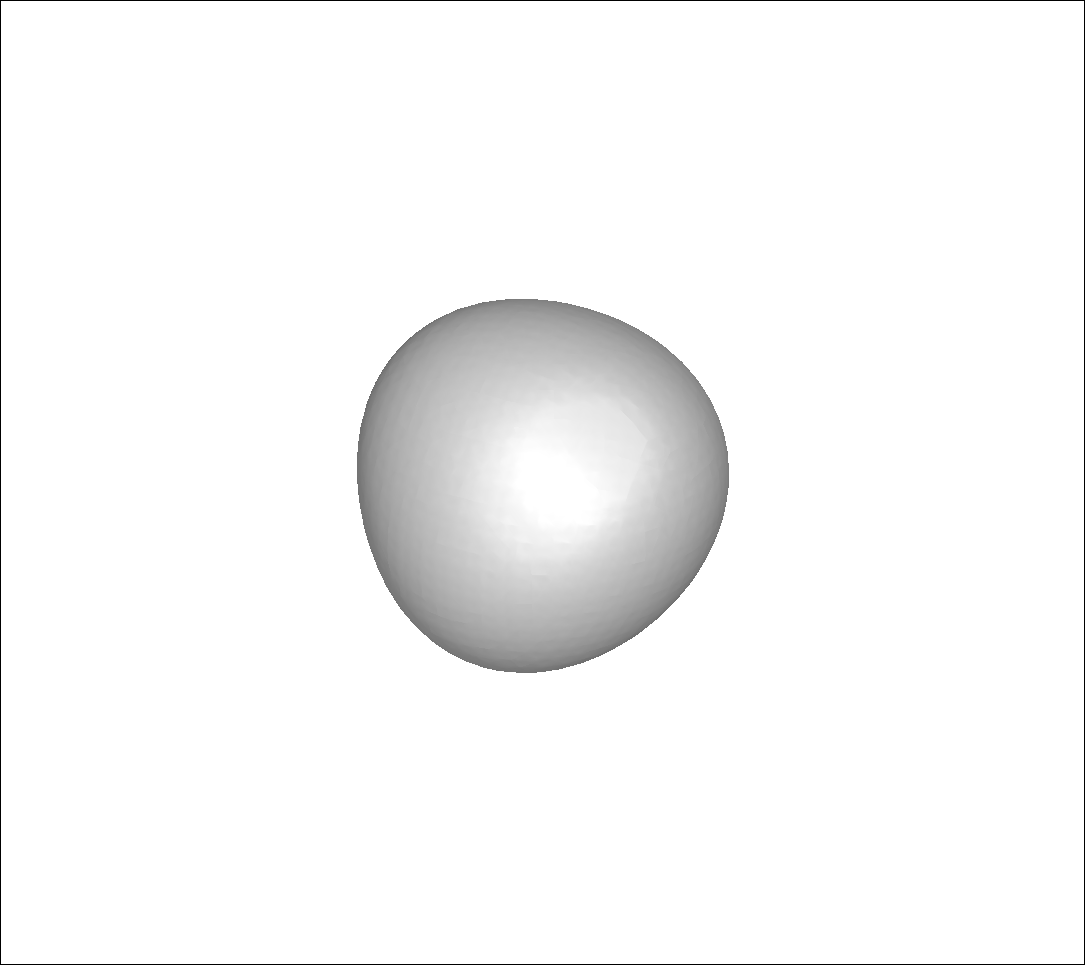} \hspace*{-0.9em}
\includegraphics[width=1.2in]{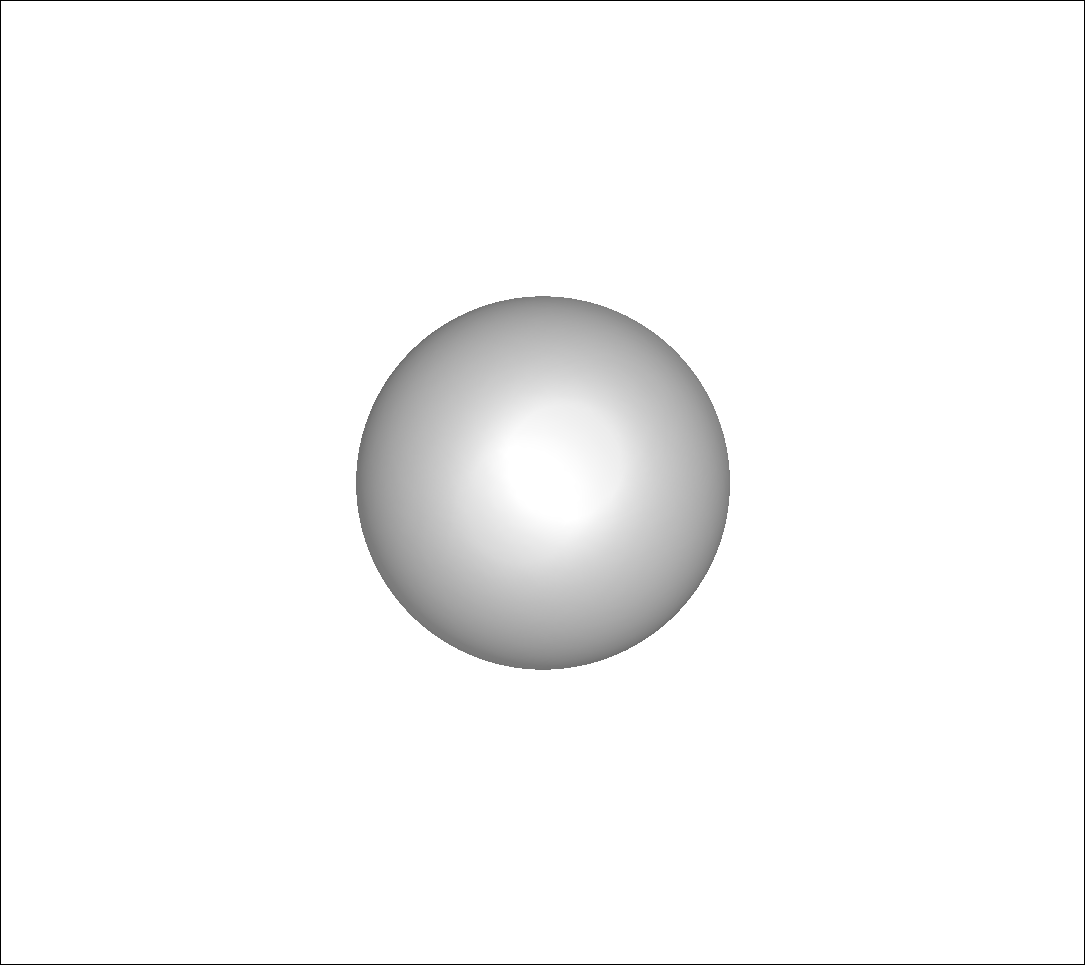}
\caption{\label{F:fig_geometries} A sequence of 3D geometries with varying curvature, but constant volume. A tetrahedron has the highest curvature while the sphere has the lowest.}
\end{center}
\end{figure*}

We use a self-consistent field theory approach~\cite{Glennbook} to model the equilibrium structures of diblock copolymer systems under 3D confinement. We utilize a finite element based (FEM) approach to numerically solve the self-consistent field theory equations \cite{ACKERMAN2017280}. A FEM based approach allows efficient computation of equilibrium structures on complex geometries with non-periodic domains. Additional advantages include the ability to enable spatial adaptivity for enhanced computational efficiency. We deploy this FEM based SCFT approach to explore equilibrium structures formed in 3D geometries with varying curvature (but having the same volume). Specifically, we consider a sequence of geometries starting with a tetrahedron and ending with a sphere, Figure.~\ref{F:fig_geometries}. This allow us to study the impact of the interplay between the curvature and confinement on the equilibrium phases in diblock copolymer systems. We explore a range of length scales as well as volume fractions and observe several interesting trends. 

\section{Methods}\label{S:methods}
The polymer system under consideration is an AB diblock copolymer melt of uniformly long chains. We use a mean field self-consistent field theory approach. The length fraction of the A block is $f_A$. The interaction between the blocks is represented in terms of the parameter, $\chi N$. The system is confined within a rigid geometry which interacts with the A and B components via an external field as described next. We explore several volume fractions, but limit simulations to one (intermediate value of) interaction parameter\footnote{The combinatorially increasing computational effort precludes realization of a complete phase diagram for each geometry for each length scale considered}, $\chi N = 18$. This $\chi N$ choice is motivated by the fact that we are not interested in the disordered phase, and all stable bulk phases in the diblock system have been observed in experimental systems at $\chi N \geq 17.67$ for varying volume fractions \citep{doi:10.1021/ma951138i}. Additionally, earlier studies of 2D diblock system under confinement \citep{Green042018} indicated that most of the stable phases in the intermediate ranges of the interaction parameter can be realized at $\chi N = 18$.

\subsection{SCFT}
We generate equilibrium microstructures of the diblock copolymer system through an iterative self-consistent field theory (SCFT) process \citep{Glennbook} using a finite element method framework.
The model and its finite element based implementation is described in detail in Ackerman et al. \cite{ACKERMAN2017280}. We briefly outline the approach, with emphasis on the addition of the interaction with the walls of the confining geometry. The Hamiltonian for the diblock system is given by: 
\begin{equation}
  \mathcal{H} = \frac{1}{V} \int d\textbf{r}(\chi N \rho _{A}(\textbf{r}) \rho _{B}(\textbf{r}) - W_{A}(\textbf{r})\rho _{A}(\textbf{r}) - W_{B}(\textbf{r})\rho _{B}(\textbf{r}) + F_{ext}(\textbf{r}) \rho _{A}(\textbf{r}) - F_{ext}(\textbf{r}) \rho _{B}(\textbf{r})) - \ln Q
\end{equation}
where $\rho _{A}$ and $\rho _{B}$ are the density fields of the A and B components, respectively; $W_{A}$ and $W_{B}$ are the potential fields of the A and B components, respectively; $V$ is the system volume; and $Q$ is the partition function for a polymer interacting with the fields. From the Hamiltonian, the SCFT equations are derived as:
\begin{align} 
  W_{A} (\textbf{r}) &= \chi N \rho _{B} (\textbf{r}) + \lambda (\textbf{r}) + F_{ext}(\textbf{r}) \label{Eq:wa} \\
  W_{B} (\textbf{r}) &= \chi N \rho _{A} (\textbf{r}) + \lambda (\textbf{r}) - F_{ext}(\textbf{r}) \label{Eq:wb} \\
  \rho _{A} (\textbf{r}) &+ \rho _{B} (\textbf{r}) = 1 \label{Eq:incompress} \\
  \rho _{A} (\textbf{r}) &= -\frac{\delta \ln Q}{\delta W_{A}} \\
  \rho _{B} (\textbf{r}) &= -\frac{\delta \ln Q}{\delta W_{B}} 
\end{align} 
where $\lambda$ is a Lagrange multiplier enforcing the incompressibility constraint (Eqn. \ref{Eq:incompress}) and $F_{ext}(\textbf{r})$ represents the force acting on the blocks by the surface of the confining geometry. We choose the force field to be attractive to block B and repulsive to block A\footnote{Mathematically, one can comfortably make this assumption without loss of generality. From an experimental standpoint this translates to functionalizing the pore material to preferentially attract one block. Computationally, this follows the approach in earlier work by Li~\cite{Li2006}.}. 

The system is solved on a mesh consisting of tetrahedral finite elements. SCFT iterations are performed until a termination criteria is satisfied (the max nodewise difference between the potential fields in successive iterations is less than a threshold value, usually $10^{-2}$). We also perform a rigorous mesh convergence analysis (please see appendix) which informs the mesh resolution for all simulations. 

\subsection{Wall-field generation} \label{S:wallfield}
To generate the wall field, we generalize a previous approach used by Green et al. \cite{Green042018} for confinement in 2D polyhedrals, to arbitrary 3D boundaries. Green et al. \cite{Green042018} extended the approach used by Li et al. \cite{Li2006} to apply to multiple confining surfaces. The wall field, $H_{wall}(\textbf{r})$, represents the interaction between a wall at the domain boundary and a polymer segment located at $\textbf{r}$:
\begin{equation} \label{eq:hwall}
  H_{wall}(\textbf{r}) =  \begin{cases}
    A_{0} \chi N [\exp (\frac{0.4R_{g}-d}{0.2R_{g}})-1], d < 0.4 R_{g} \\
    0, \ \ \ \ \ \ \ d \geq 0.4 R_{g}
    \end{cases}  
\end{equation}
where $A_{0} =0.4$, $d$ is the distance between the wall and the polymer segment located at \textbf{r}. Li et al. \citep{Li2006} note that modest variations in the strength of the surface field (with $A_{0}$ in the range, $0.1 < A_{0} < 0.4$) did not alter the morphologies significantly. The full external field at location $\textbf{r}$ is the sum of the fields from all of the walls:
\begin{equation} \label{eq:fwall2d}
  F_{ext}(\textbf{r}) = \sum _{i} H_{wall}^{i}(\textbf{r})
\end{equation}
This method is suitable for the case of regular polyhedrals with a finite number of walls, but it is less suitable for an arbitrary geometry. To extend this method to confinement in an arbitrary 3D geometry, we compute a surface integral taking into account the contribution to the force field from each point on the surface (essentially taking the summation in Eqn.~\ref{eq:fwall2d} to its integral limit). The interaction in the integral is taken to be of the same form as Eqn.~\ref{eq:hwall}. Accordingly, the wall field is given by:

\begin{equation} \label{eq:fwall3d}
  F_{ext}(\textbf{r}) = \begin{cases}
    A_{wall} \bigintssss_{S}d \Omega \ \chi N \ [\exp (\frac{0.4R_{g}- |\textbf{r} - \textbf{s}|}{0.2R_{g}})-1],  |\textbf{r} - \textbf{s}| < 0.4 R_{g} \\
    0, \ \ \ \ \ \ \ |\textbf{r} - \textbf{s}| \geq 0.4 R_{g}
    \end{cases}
\end{equation}

where $\textbf{s}$ represents a location on the surface domain, $\Omega$ and $A_{wall}$ is a normalization constant. The same $A_{wall}$ is used for wall field calculation in geometries with different curvature and volume. This ensures a consistent comparison of the results, and is experimentally meaningful. The calculation of the normalization constant, $A_{wall}$ is further described in \ref{S:Awall}.

\subsection{Generation of 3D geometries}
We generate a series of 3D geometries smoothly changing from a tetrahedron to a sphere. We use a curvature-flow evolution solver to construct these geometries~\citep{WODO20116037}. This ensures that the volume of the geometries all remain identical. We analyze a total of 5 geometries including the tetrahedron, sphere, and 3 intermediate geometries. The outlines of the geometries are shown below in Figure~\ref{F:fig_geometries}. We define a baseline geometry size as the tetrahedron with edge length $L = 14.72 R_{g}$. As the geometry changes from a tetrahedron to sphere through curvature driven flow (at constant volume), the maximum distance of the surface to the center decreases monotonically. Therefore, each geometry can be uniquely identified by this distance. For the baseline geometry size, the maximum distances from the surface of the five geometries considered here to the center are $9.01 R_{g}$,  $7.11 R_{g}$,  $5.64 R_{g}$,  $4.78 R_{g}$  and $4.48 R_{g}$ respectively.

\section{Results and discussion} \label{S:results}

Using the methods and geometries above, we first look at the equilibrium microstructures for a diblock copolymer system with $f_{A} = 30$. For the case of a tetrahedron with baseline geometry size (edge length $L = 14.72 R_{g}$), the structure of the B component is fully connected from the outside to the interior for the tetrahedron geometry while it is disconnected in the spherical geometry. We explore this transition of the equilibrium morphologies from interpenetrating structures to non-interpenetrating structures as a function of curvature and volume fraction. We also investigate the effect of varying confinement volume on the equilibrium microstructures. To do this, we select four confinement sizes with volumes corresponding to tetrahedrons with edge lengths equal to $\frac{L}{4}$, $\frac{L}{2}$, $L$ and $2L$ where $L = 14.72 R_{g}$. We denote the volumes of tetrahedron corresponding to the edge lengths as $V_{1}$, $V_{2}$, $V_{3}$, $V_{4}$ respectively. For each of the five geometries considered here (with varying curvature), we explore the equilibrium microstructures with confinement volumes, $V_{1}$, $V_{2}$, $V_{3}$ and $V_{4}$.

\subsection{Variation of curvature and confinement volumes with fixed volume fraction, $f_{A}=30$}

\begin{figure*}
\begin{center}
\includegraphics[width=1.0in,height=0.9in]{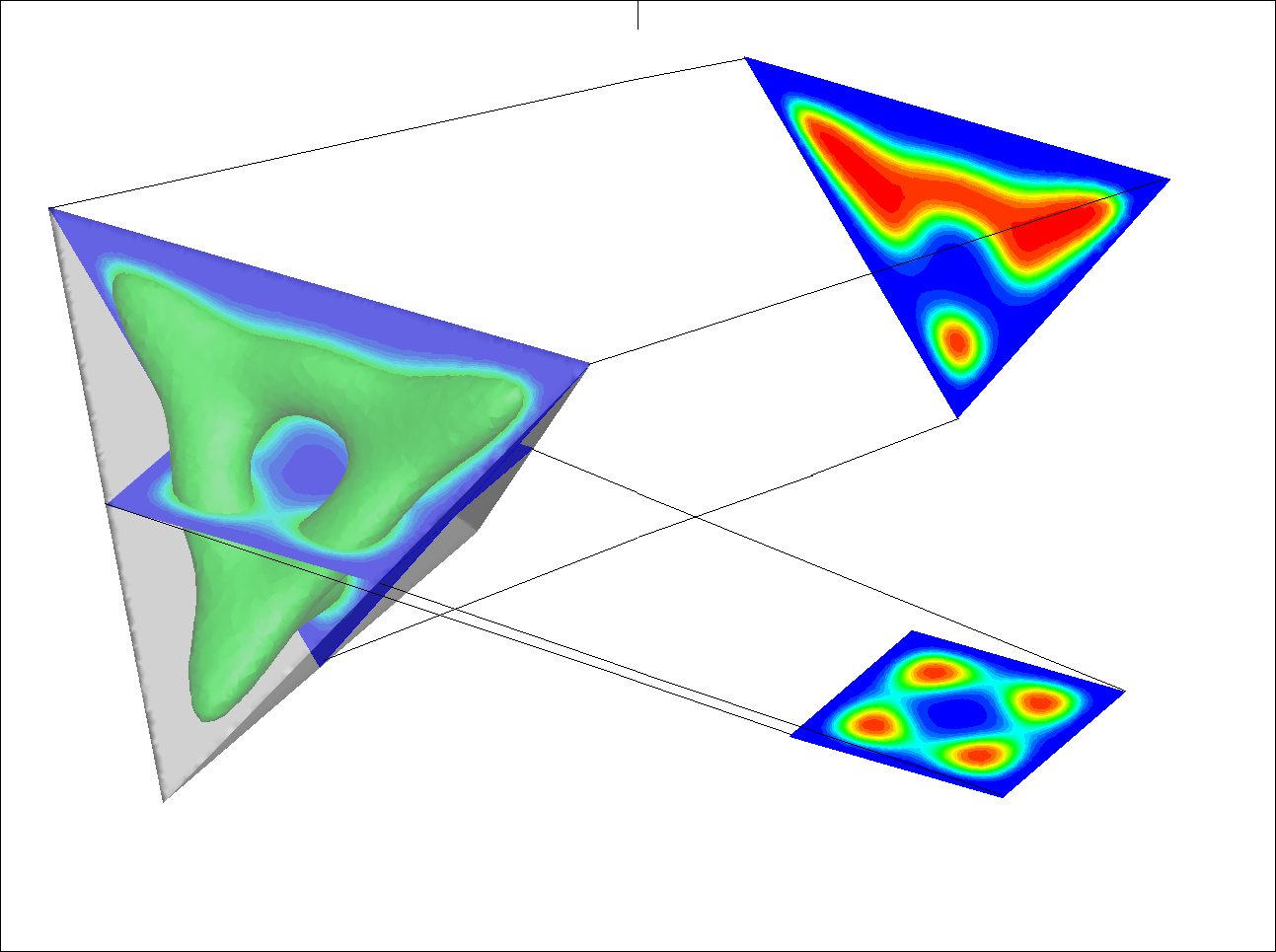} \hspace*{-0.9em}
\includegraphics[width=1.2in,height=0.9in]{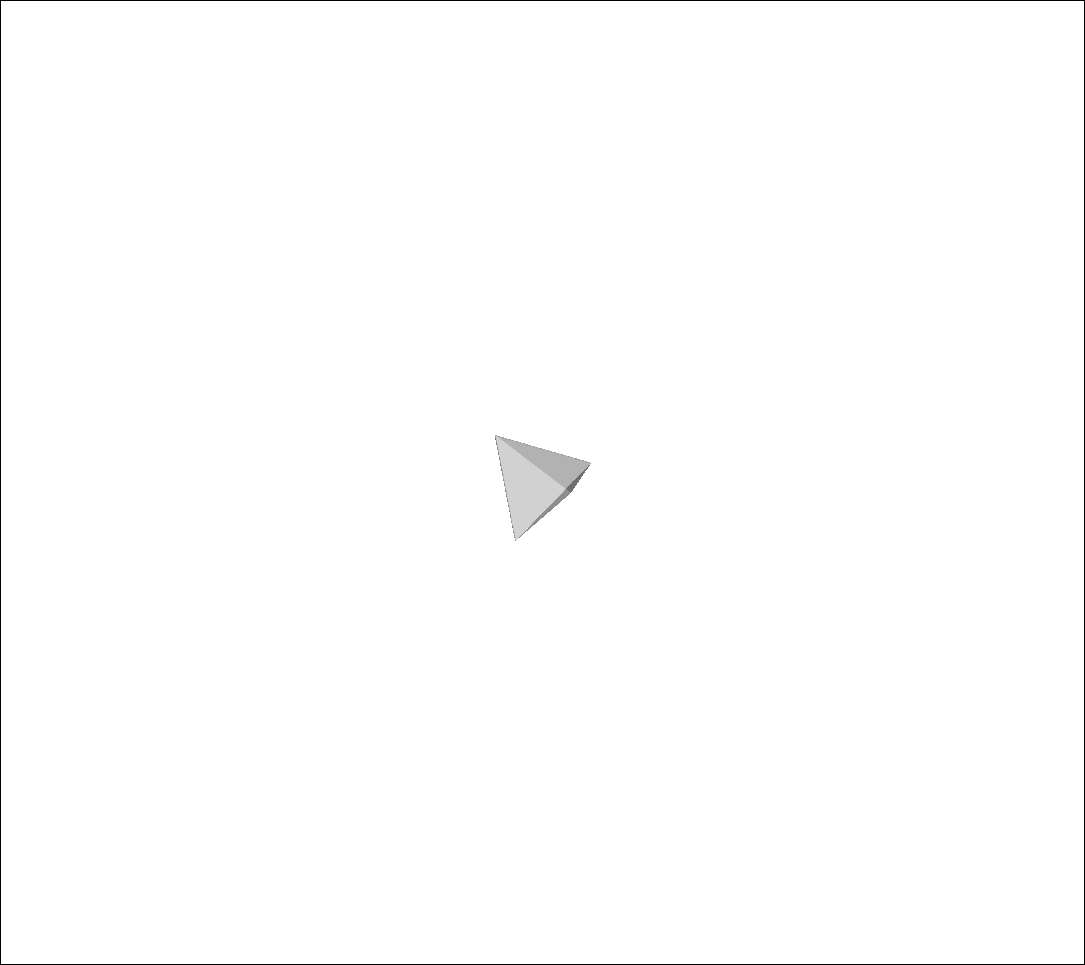} \hspace*{-0.9em}
\includegraphics[width=1.2in,height=0.9in]{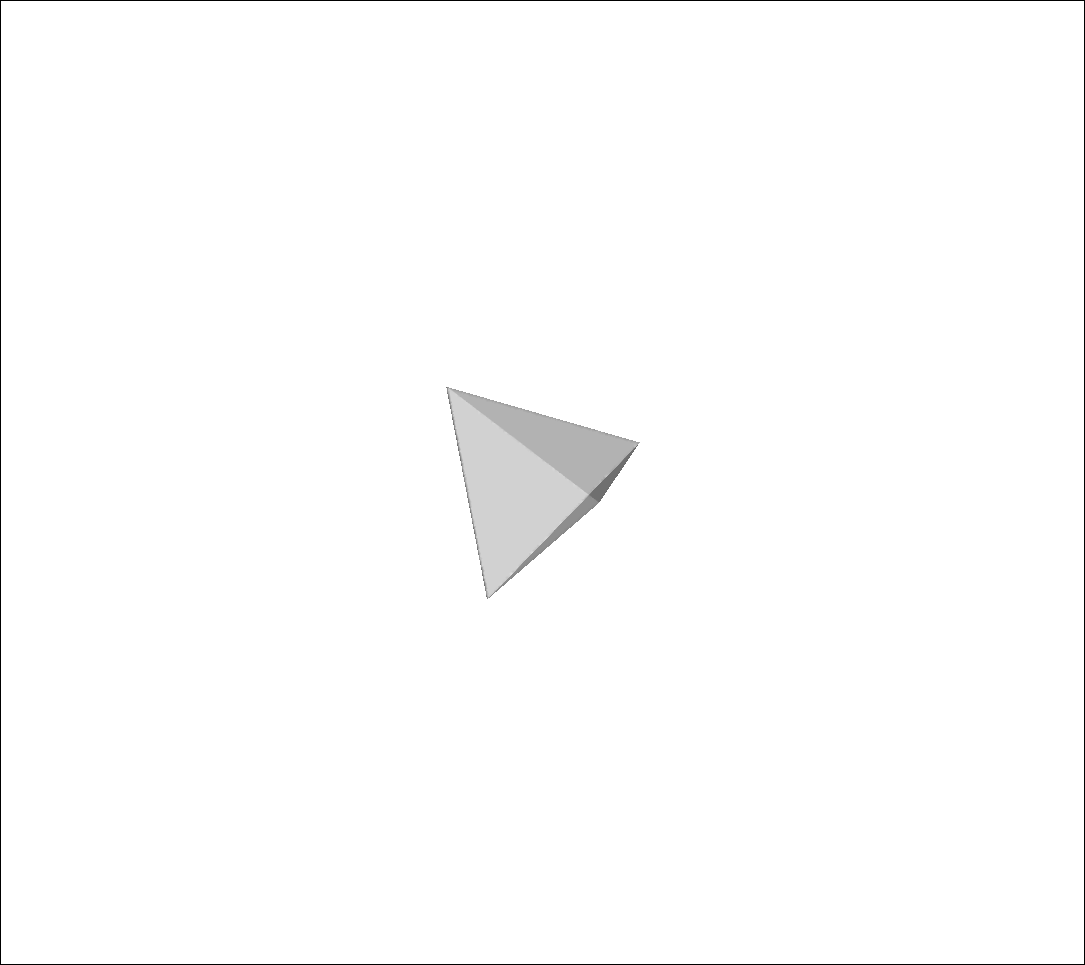} \hspace*{-0.9em}
\includegraphics[width=1.2in,height=0.9in]{tet_d1} \hspace*{-0.9em}
\includegraphics[width=1.2in,height=0.9in]{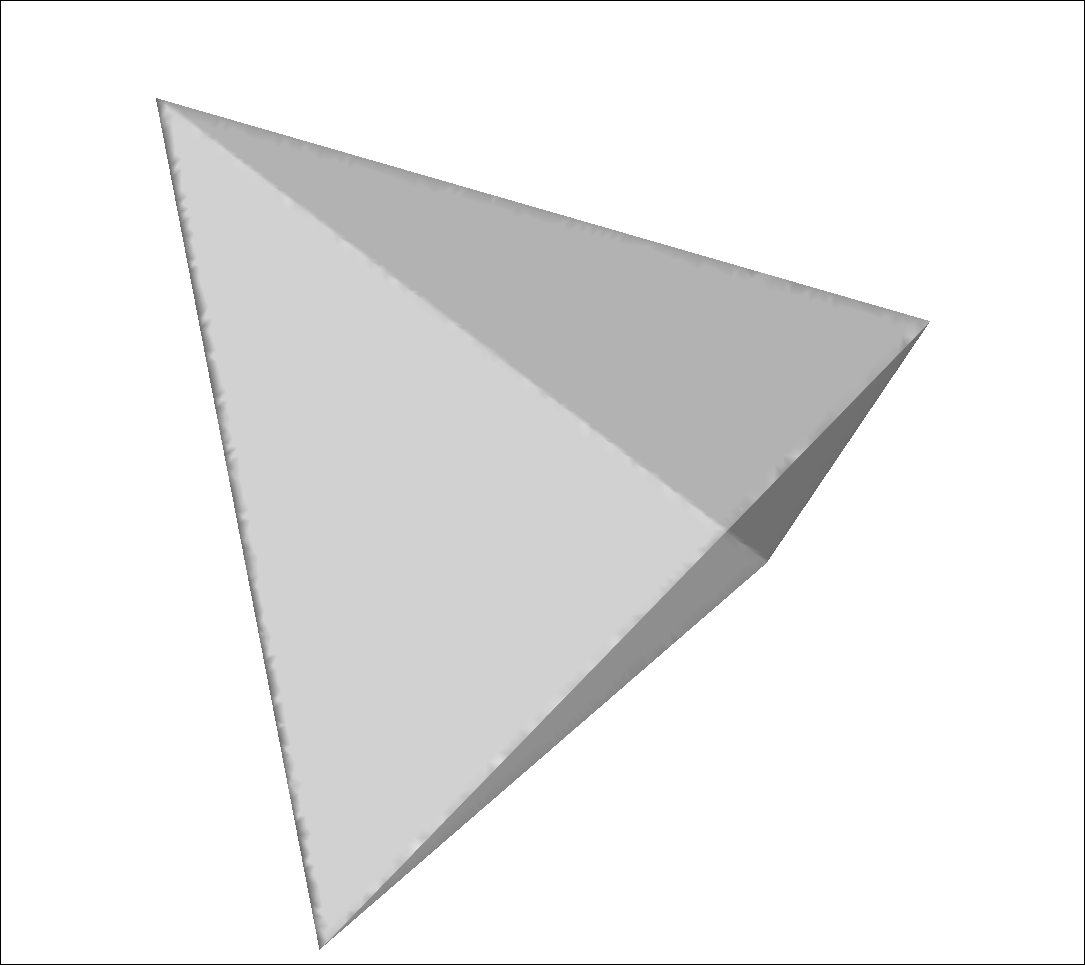}\\
\vspace{-0.08em}
\includegraphics[width=1.0in]{tet_d1} \hspace*{-0.9em}
\includegraphics[width=1.2in]{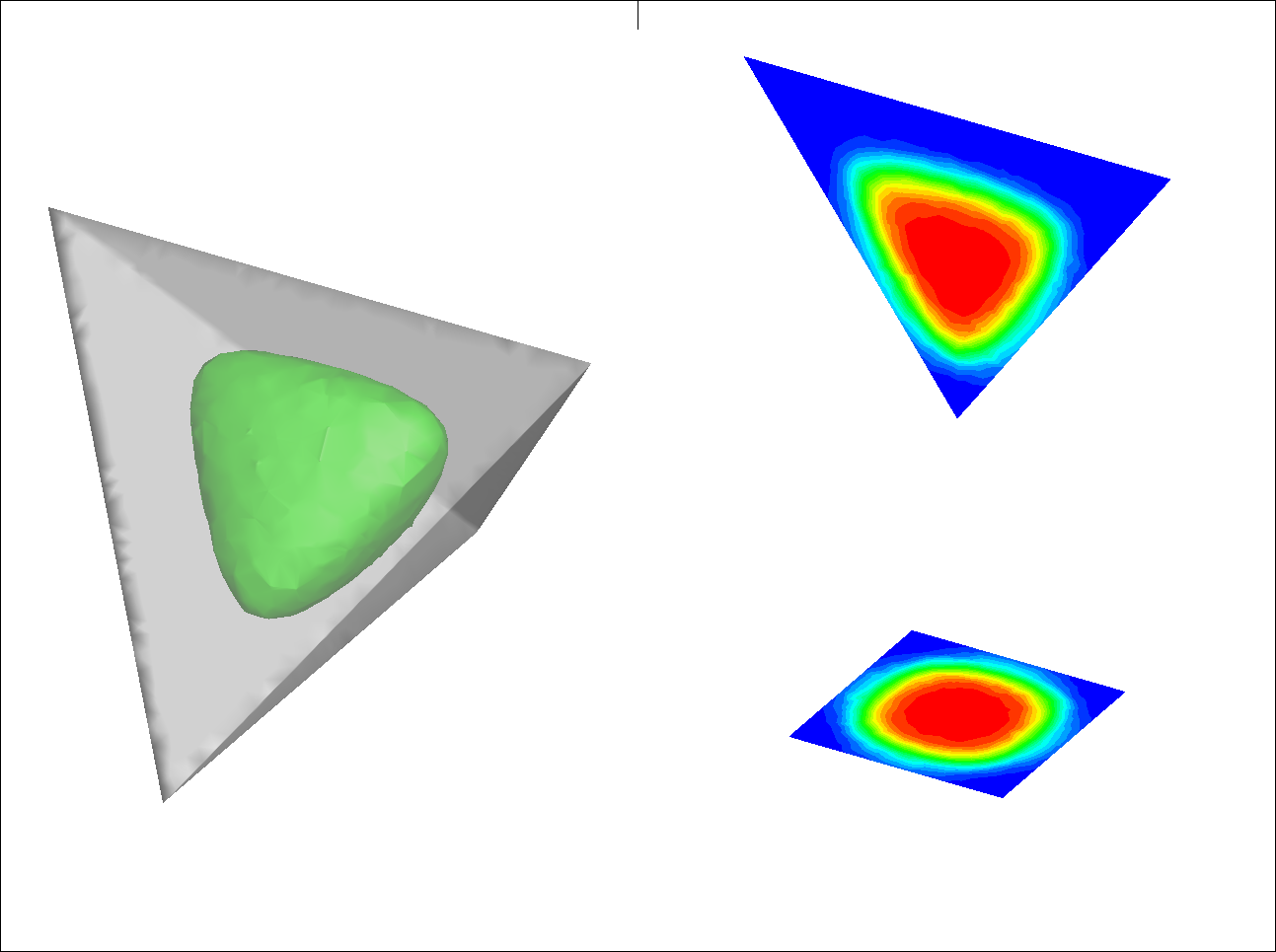} \hspace*{-0.9em}
\includegraphics[width=1.2in]{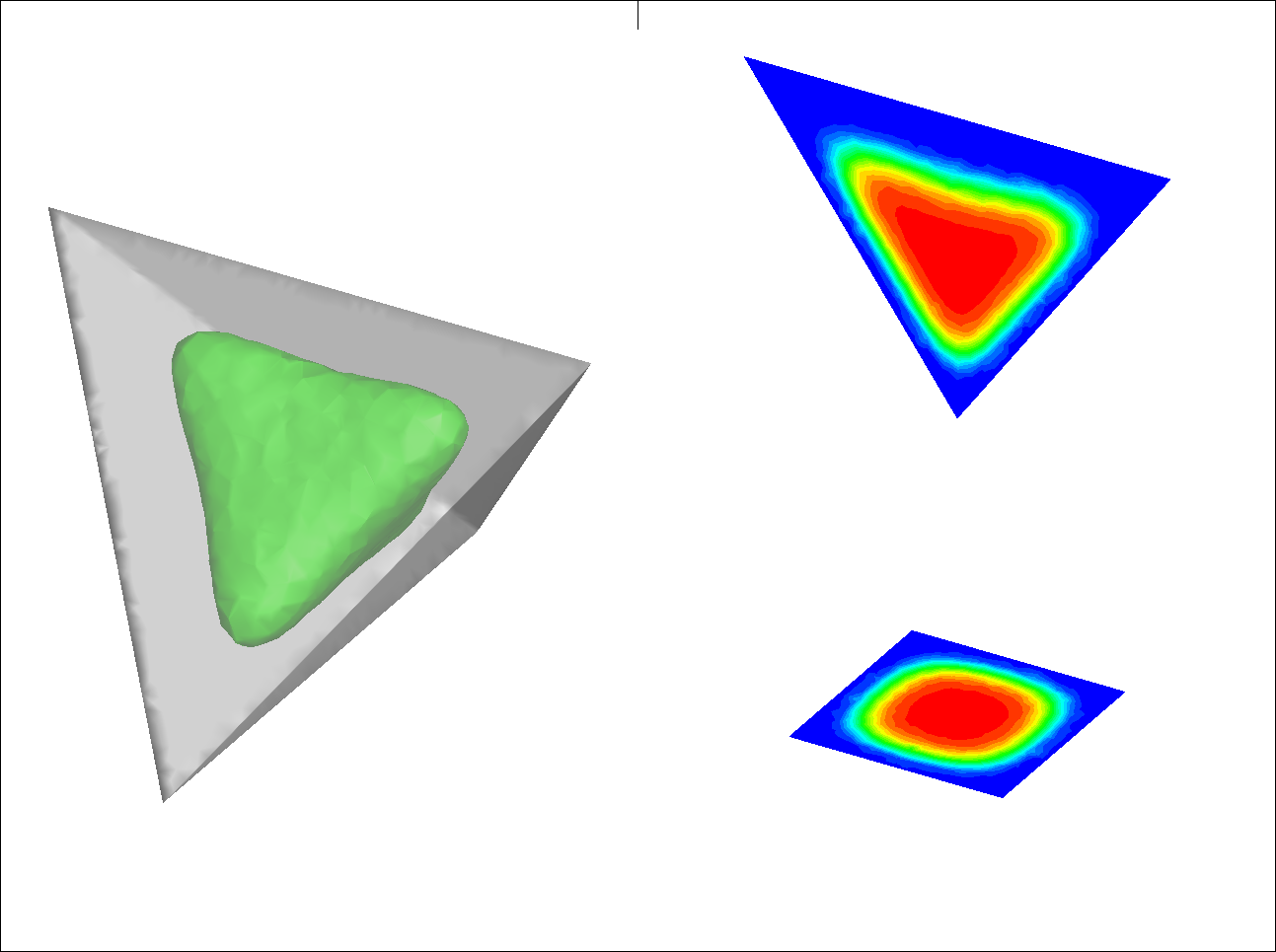} \hspace*{-0.9em}
\includegraphics[width=1.2in]{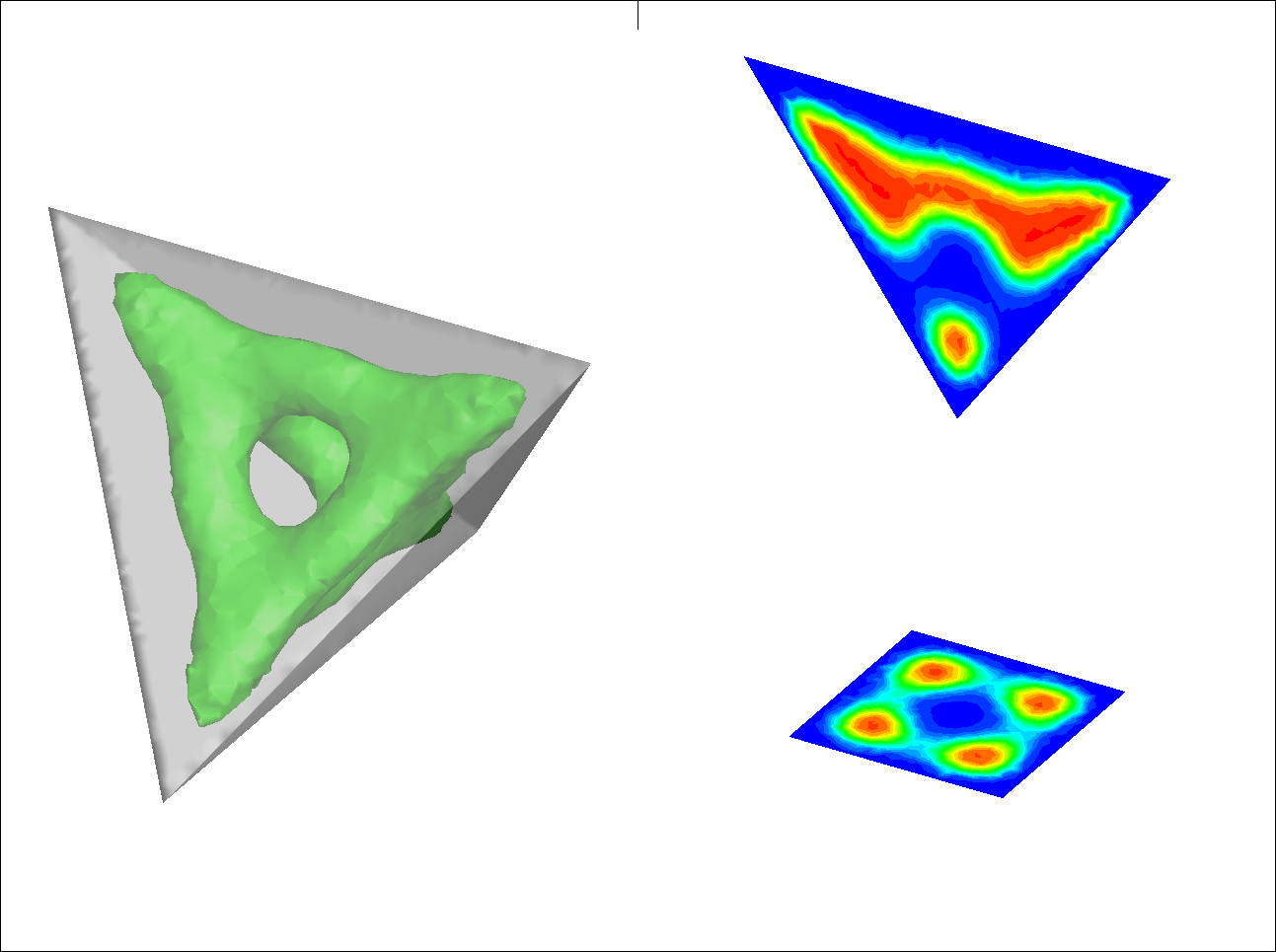} \hspace*{-0.9em}
\includegraphics[width=1.2in]{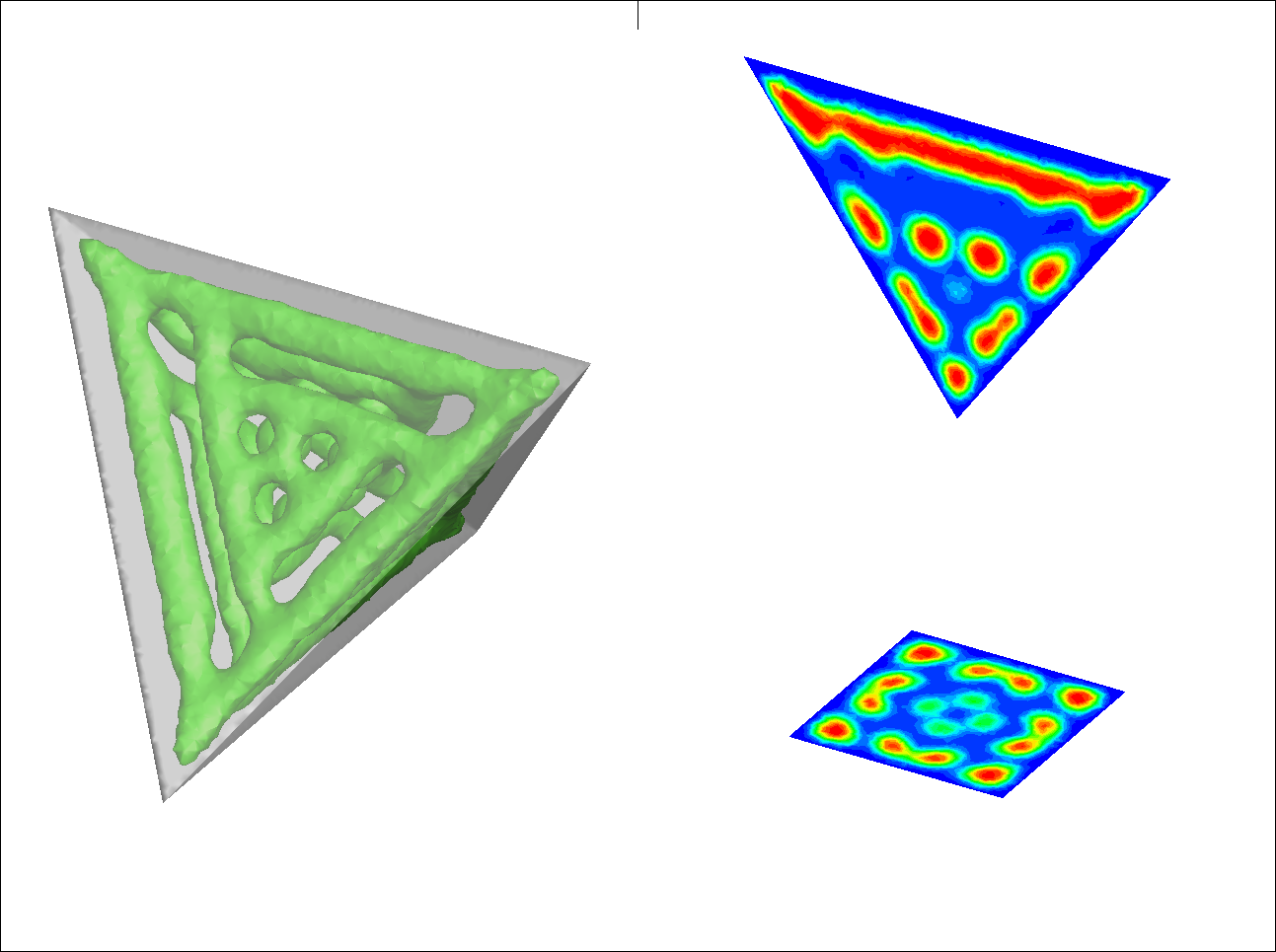}\\
\vspace{-0.08em}
\includegraphics[width=1.0in]{curv0_d1} \hspace*{-0.9em}
\includegraphics[width=1.2in]{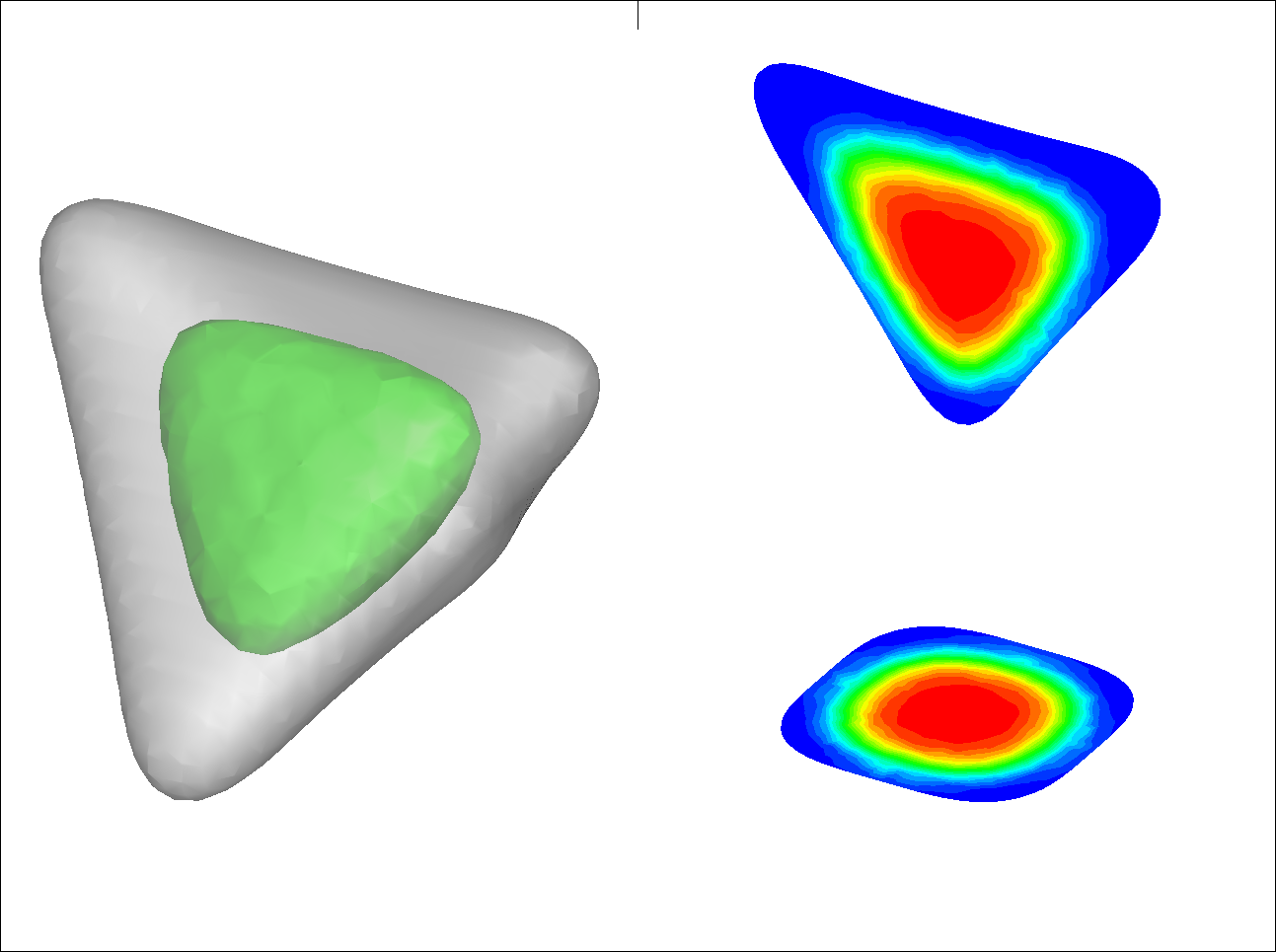} \hspace*{-0.9em}
\includegraphics[width=1.2in]{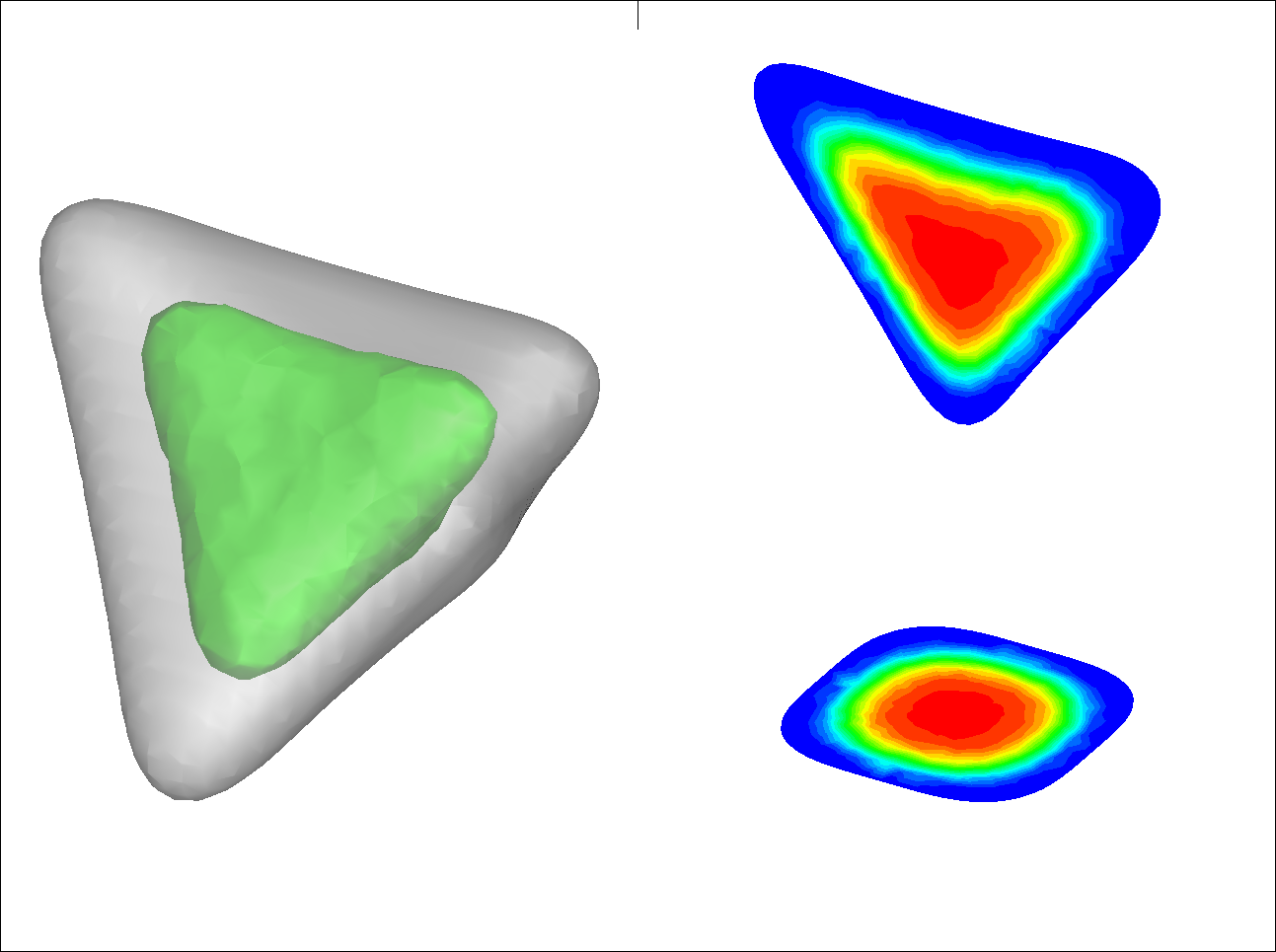} \hspace*{-0.9em}
\includegraphics[width=1.2in]{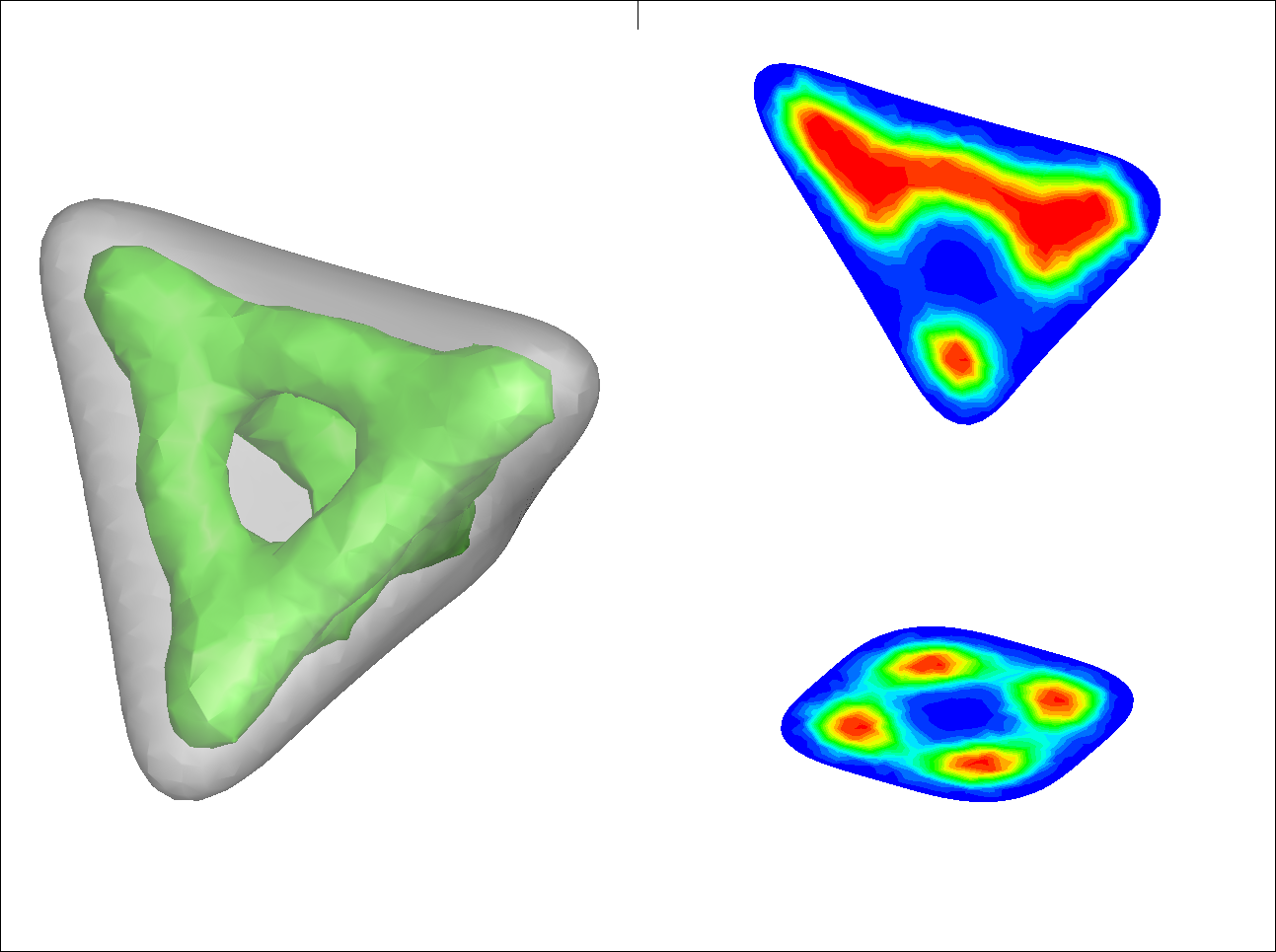} \hspace*{-0.9em}
\includegraphics[width=1.2in]{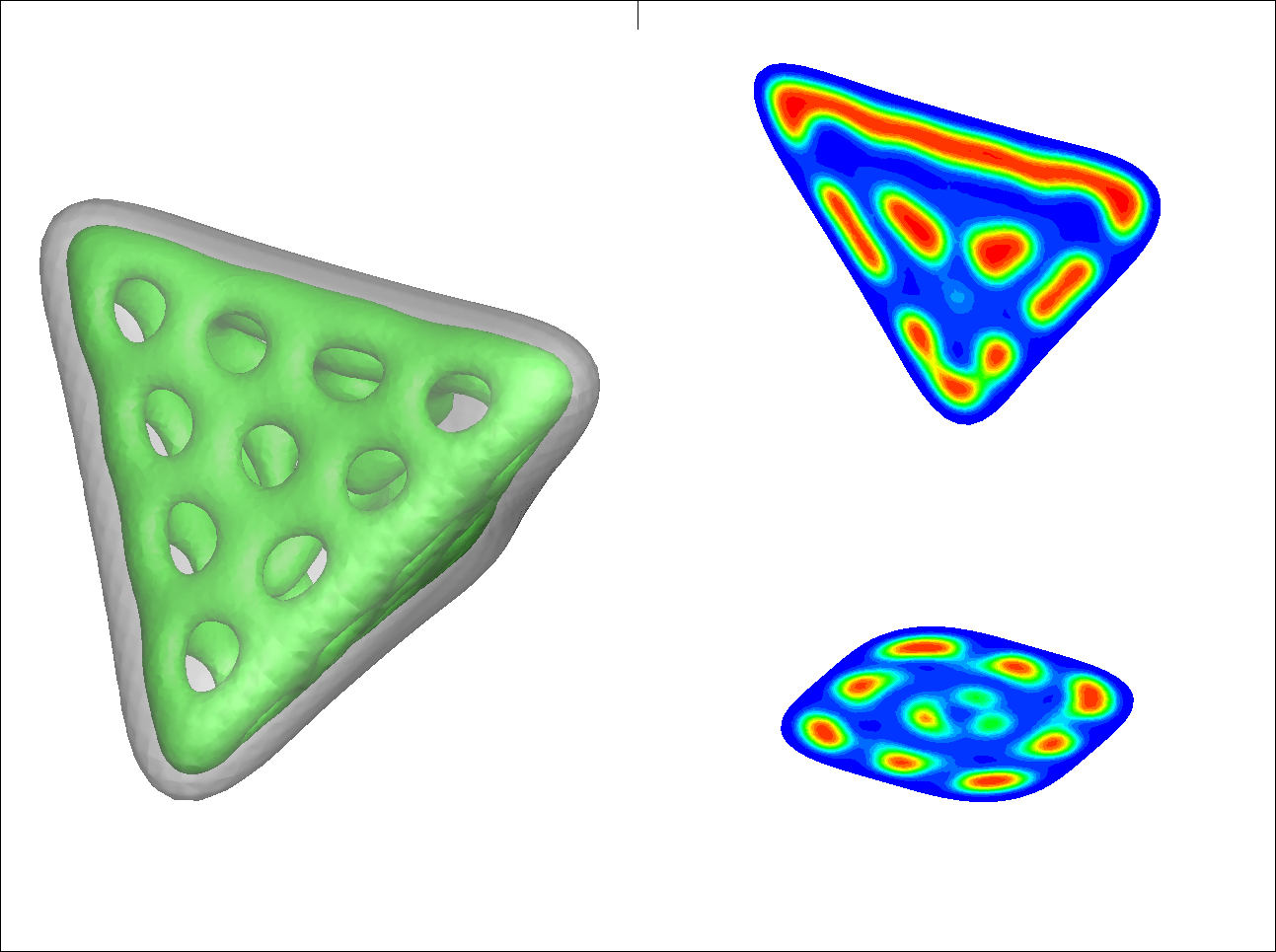}\\
\vspace{-0.08em}
\includegraphics[width=1.0in]{curv6_d1} \hspace*{-0.9em}
\includegraphics[width=1.2in]{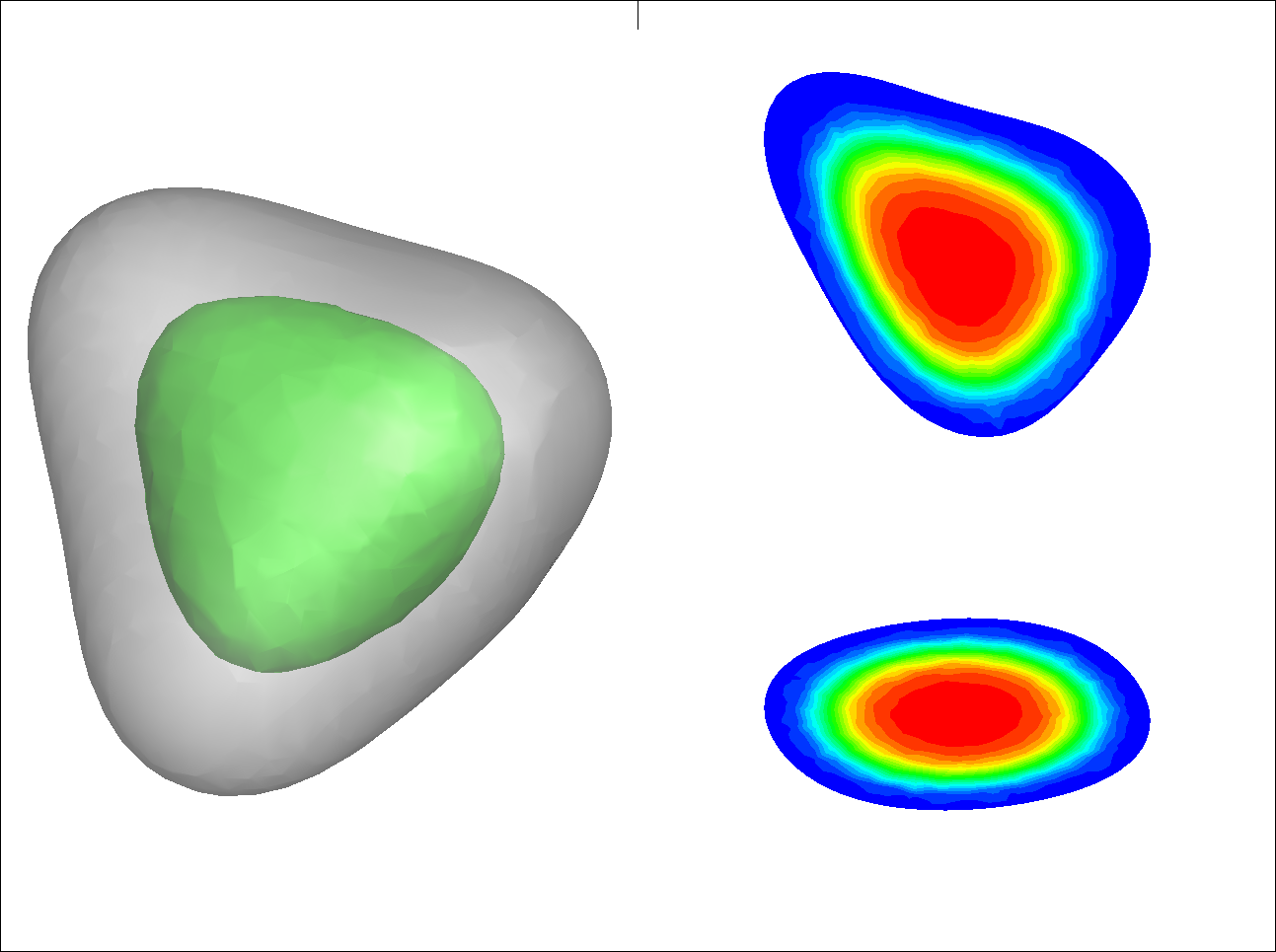} \hspace*{-0.9em}
\includegraphics[width=1.2in]{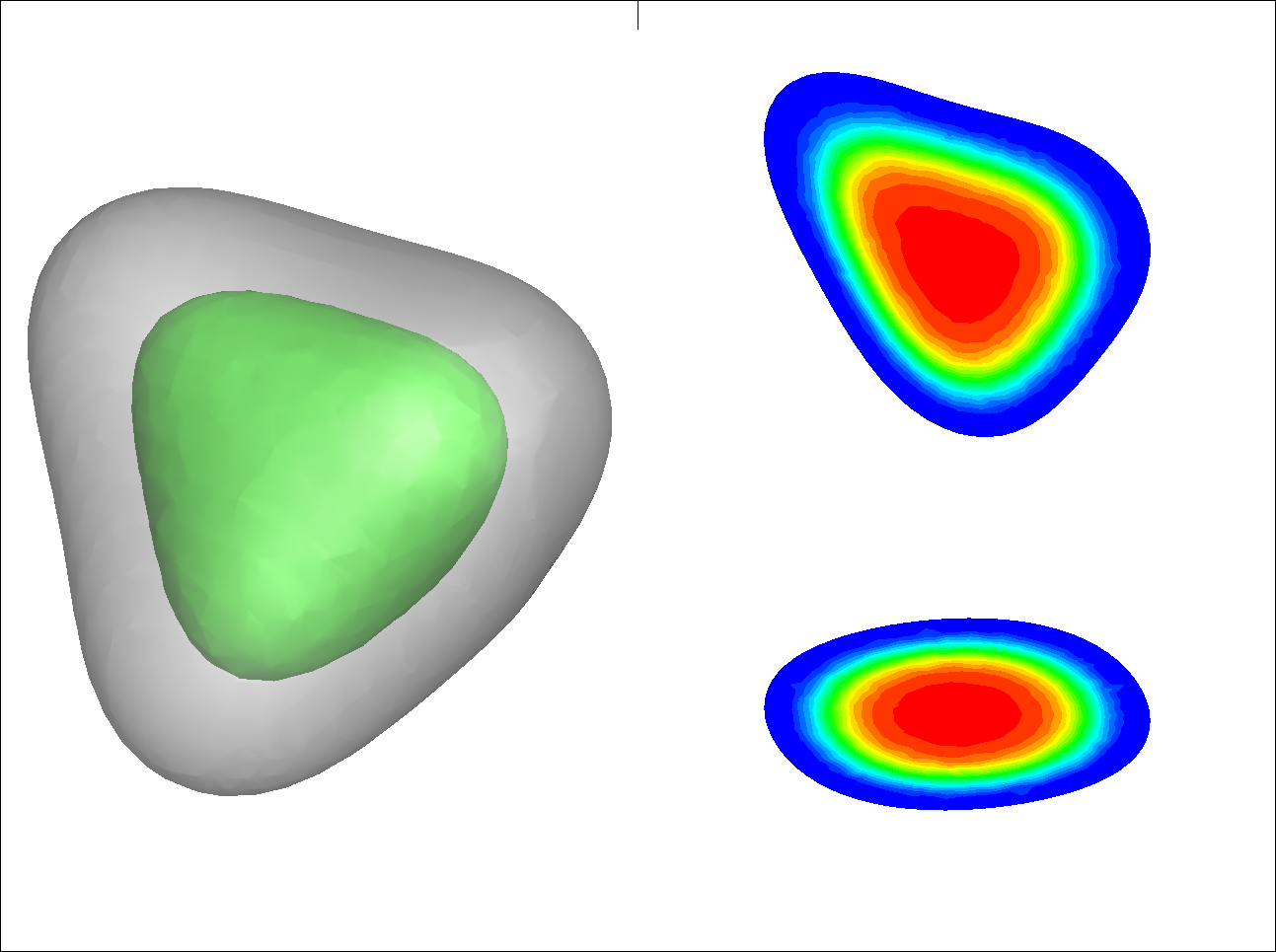} \hspace*{-0.9em}
\includegraphics[width=1.2in]{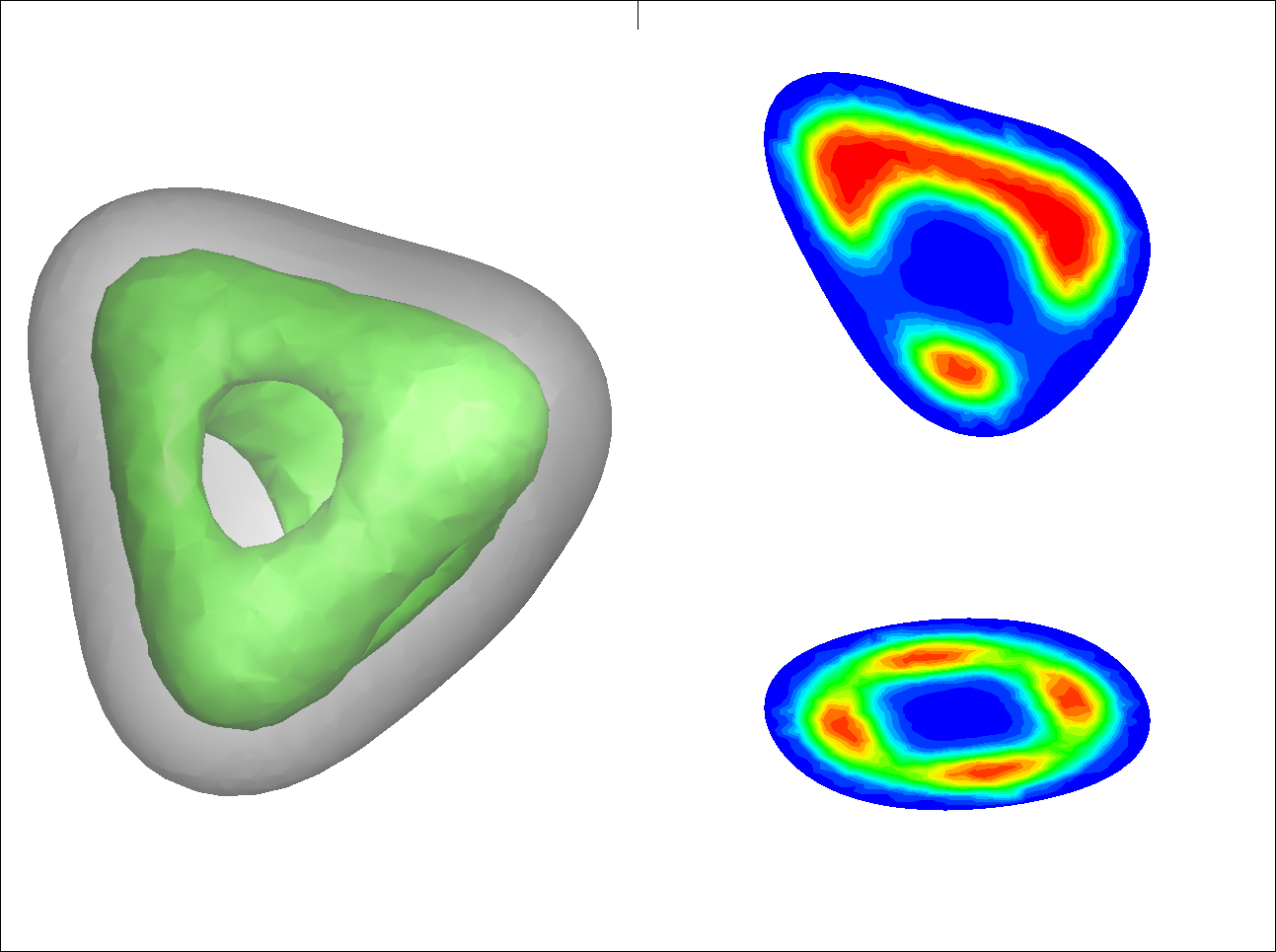} \hspace*{-0.9em}
\includegraphics[width=1.2in]{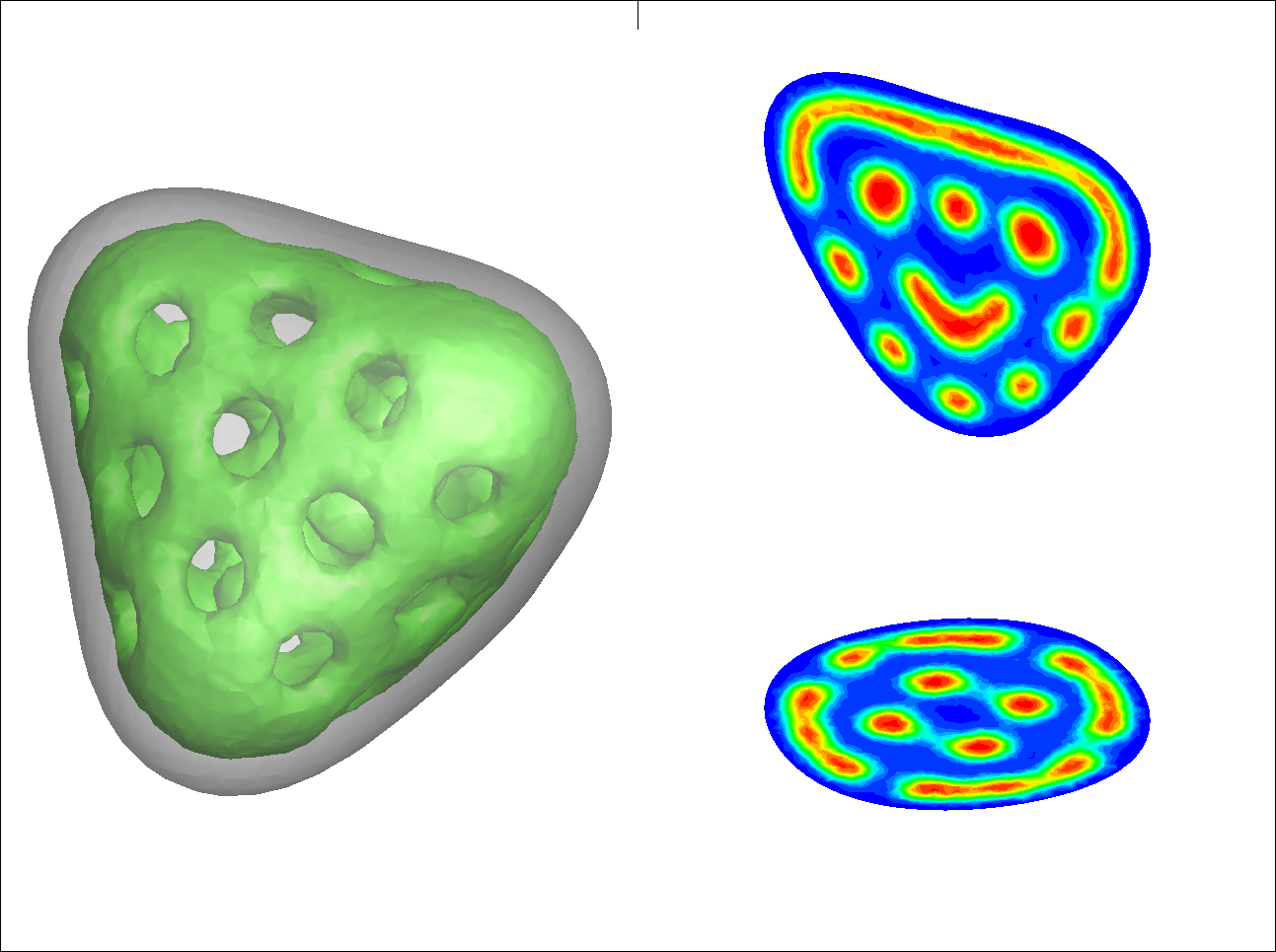}\\
\vspace{-0.08em}
\includegraphics[width=1.0in]{curv16_d1} \hspace*{-0.9em}
\includegraphics[width=1.2in]{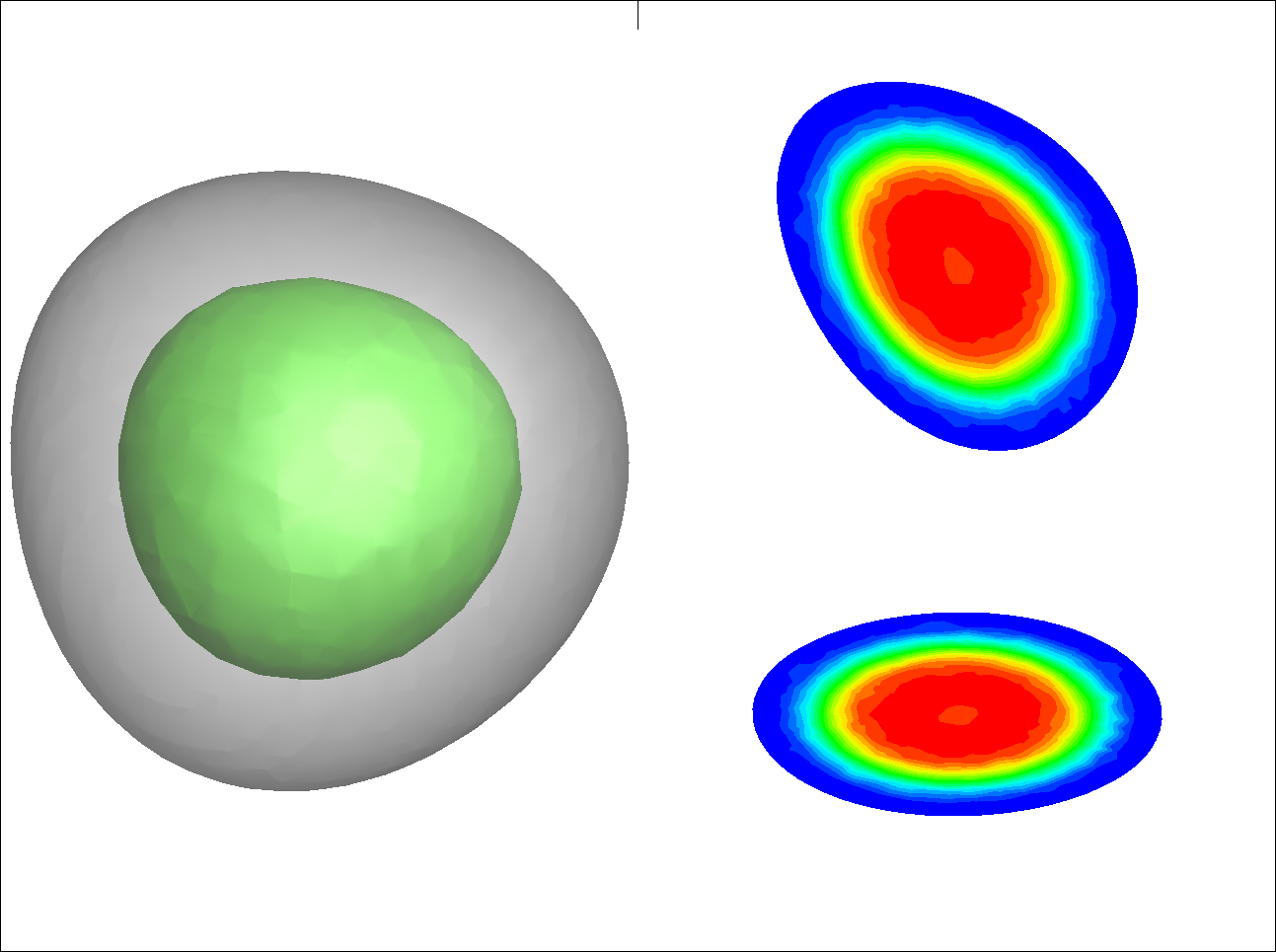} \hspace*{-0.9em}
\includegraphics[width=1.2in]{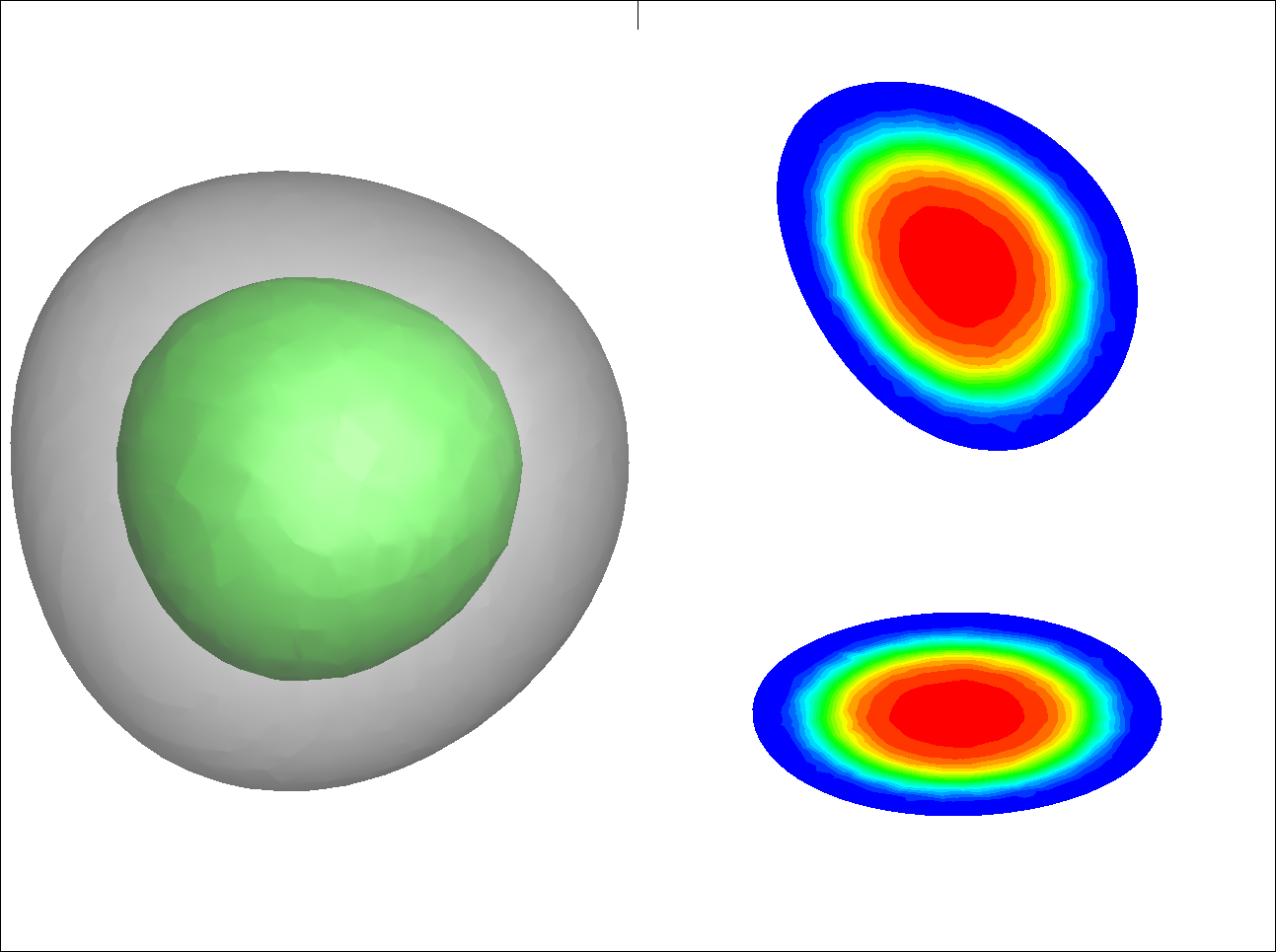} \hspace*{-0.9em}
\includegraphics[width=1.2in]{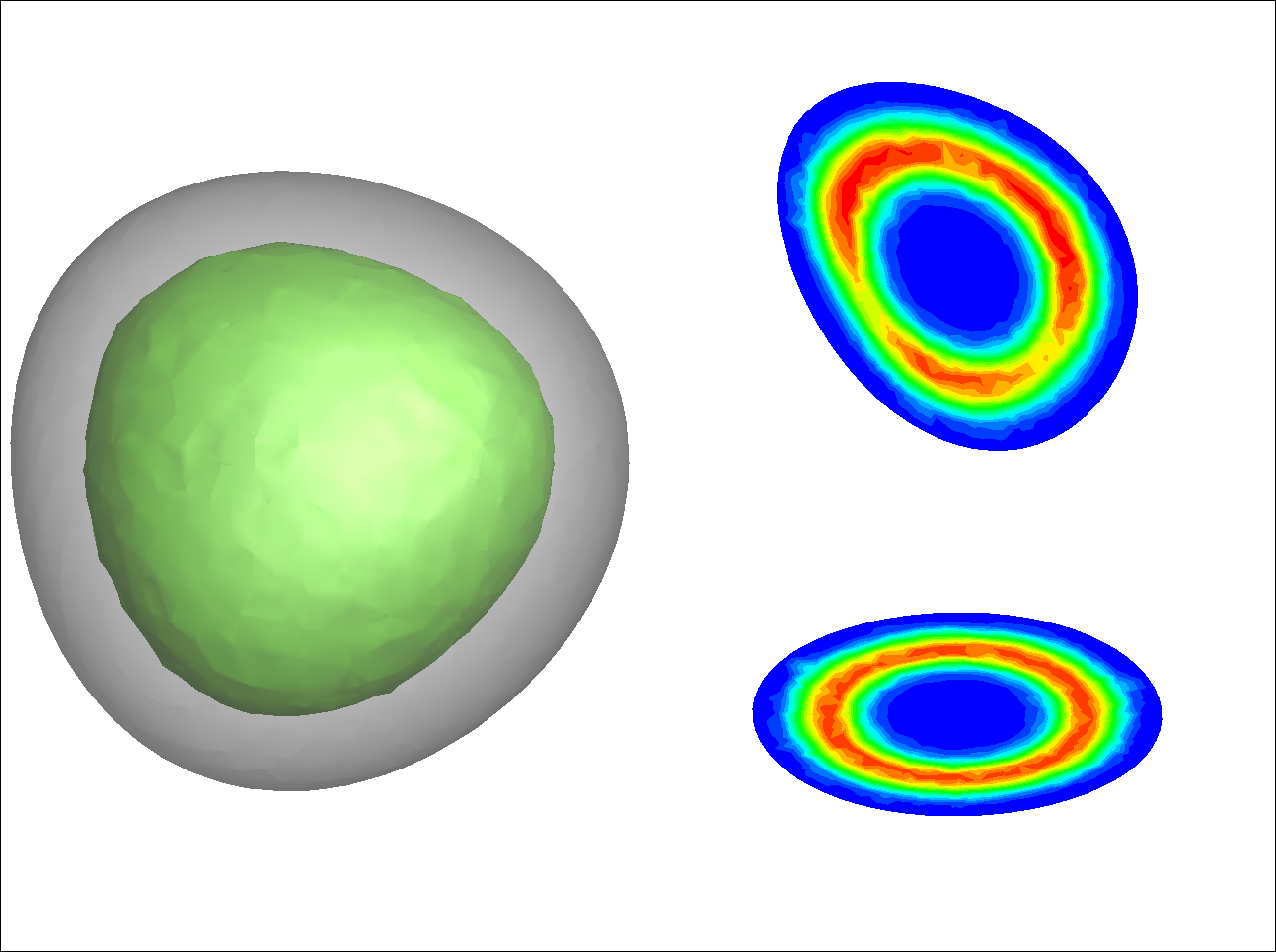} \hspace*{-0.9em}
\includegraphics[width=1.2in]{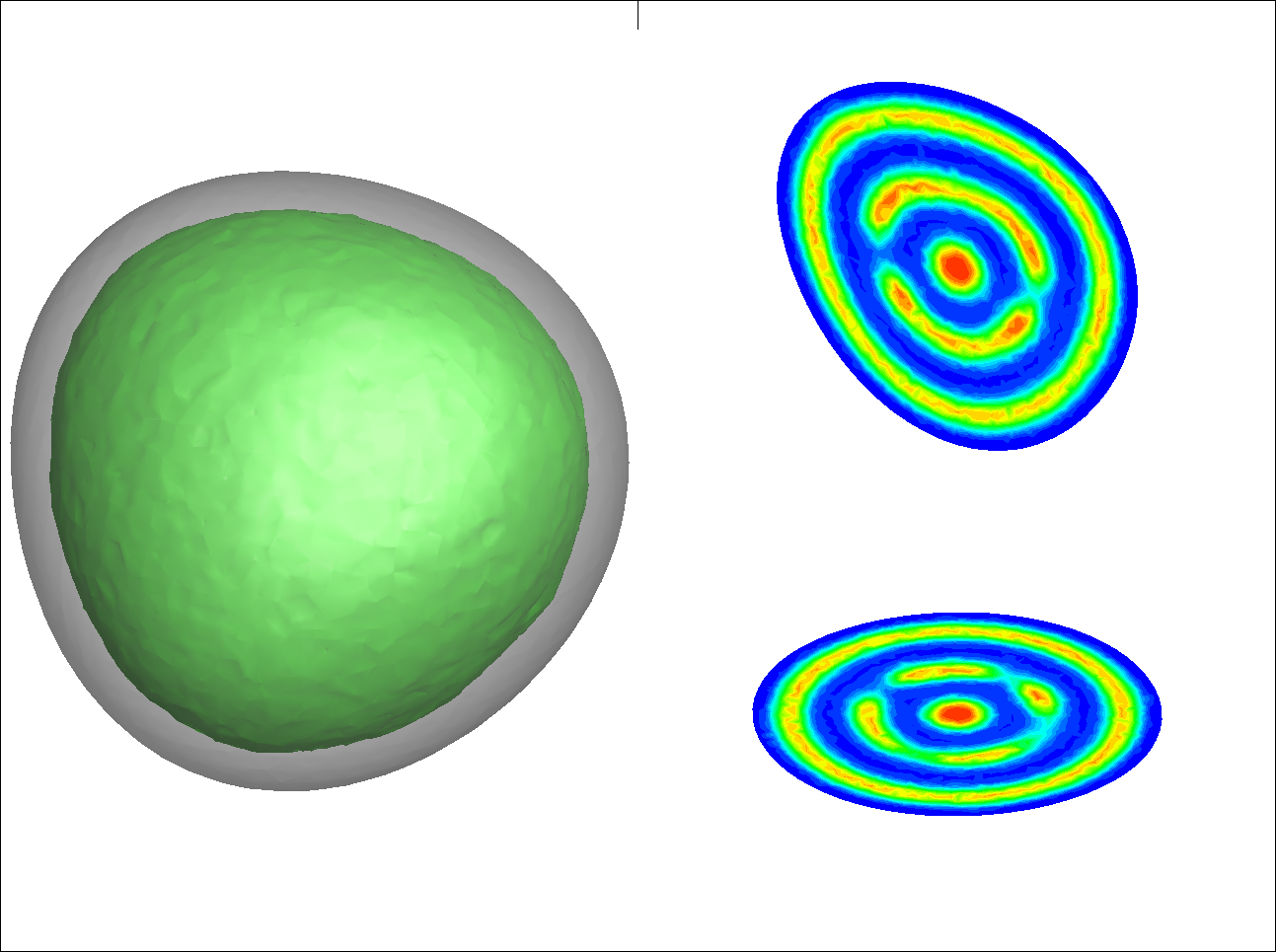}\\
\vspace{-0.08em}
\includegraphics[width=1.0in]{sphere_d1} \hspace*{-0.9em}
\includegraphics[width=1.2in]{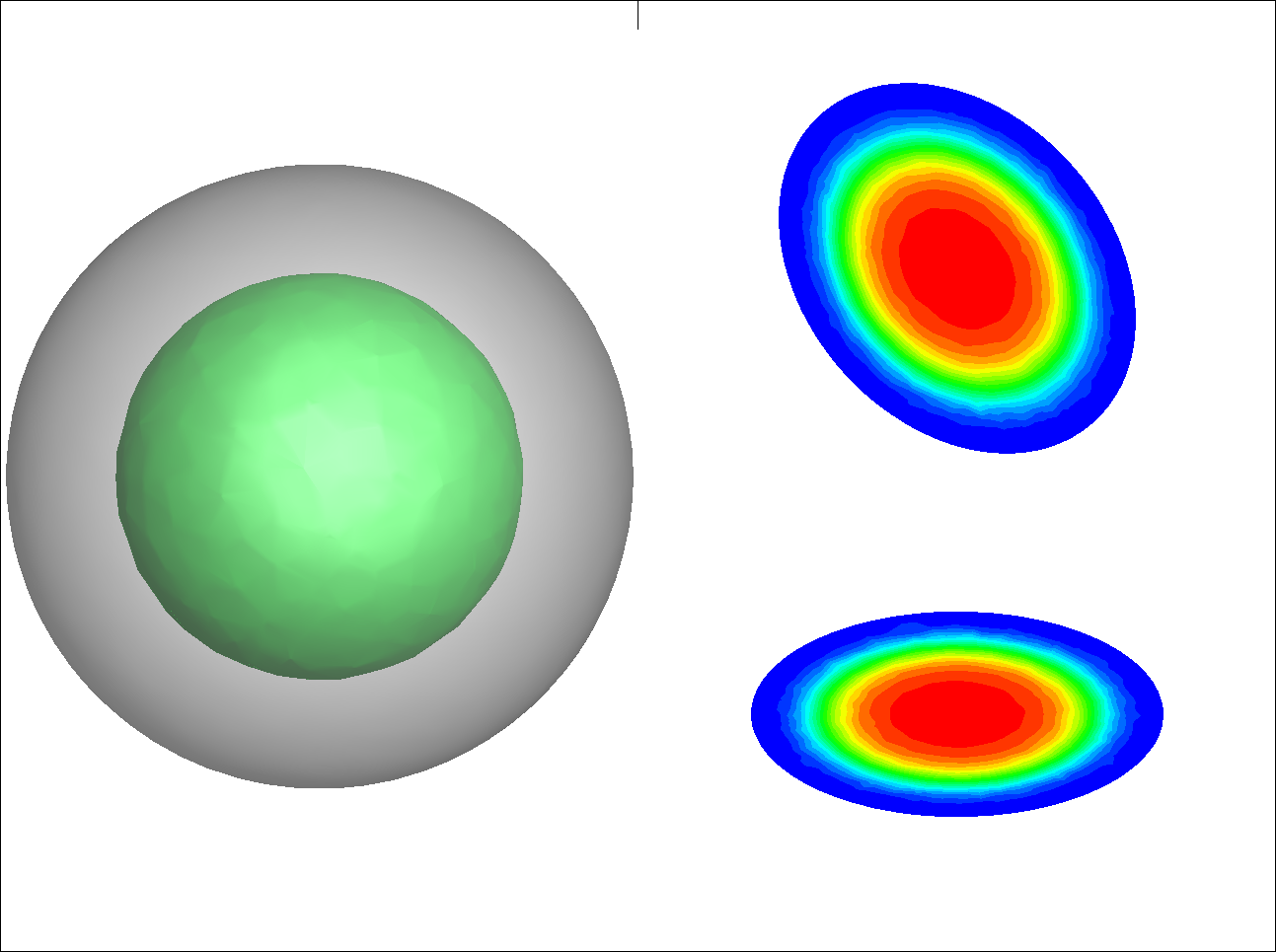} \hspace*{-0.9em}
\includegraphics[width=1.2in]{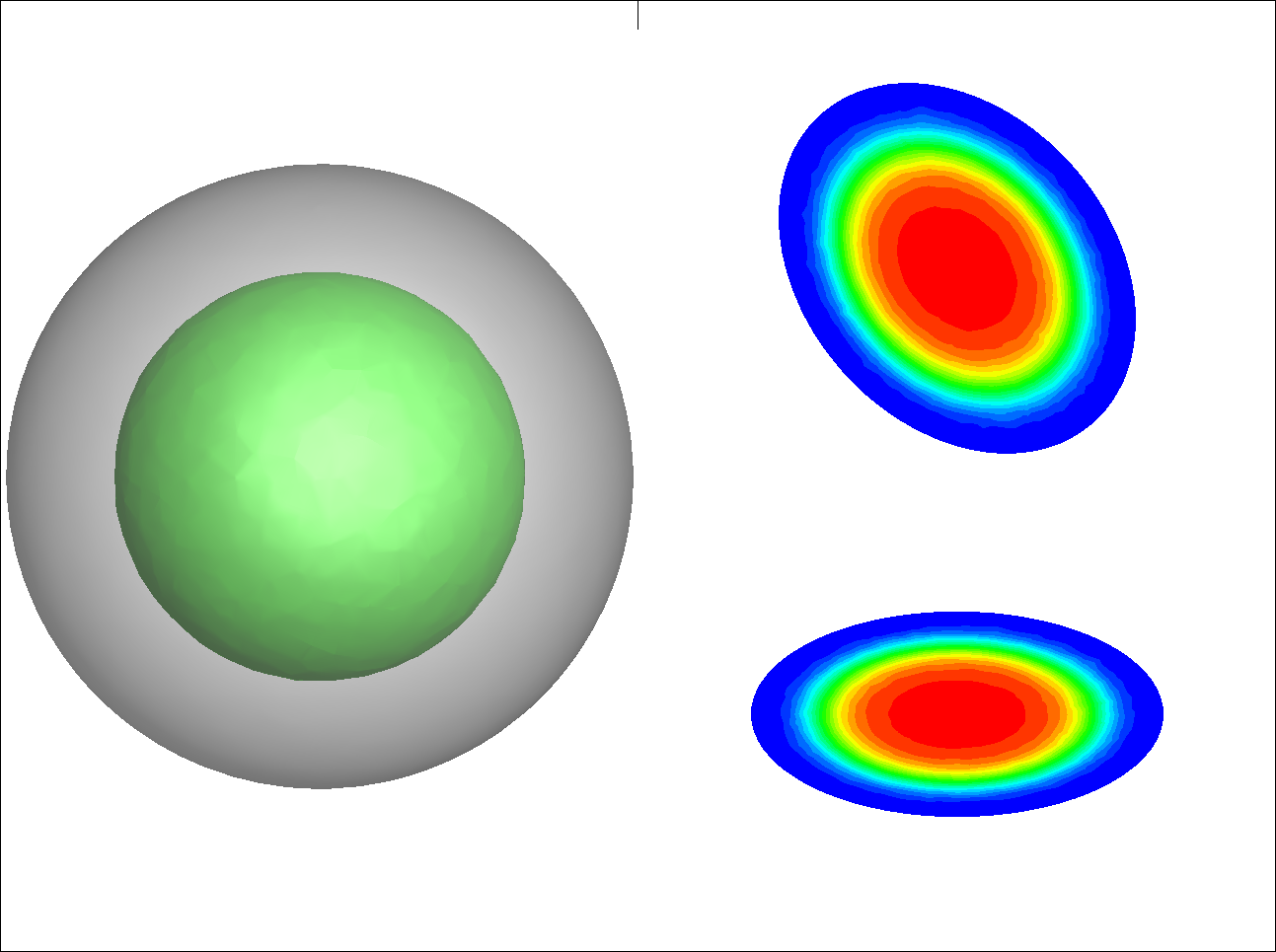} \hspace*{-0.9em}
\includegraphics[width=1.2in]{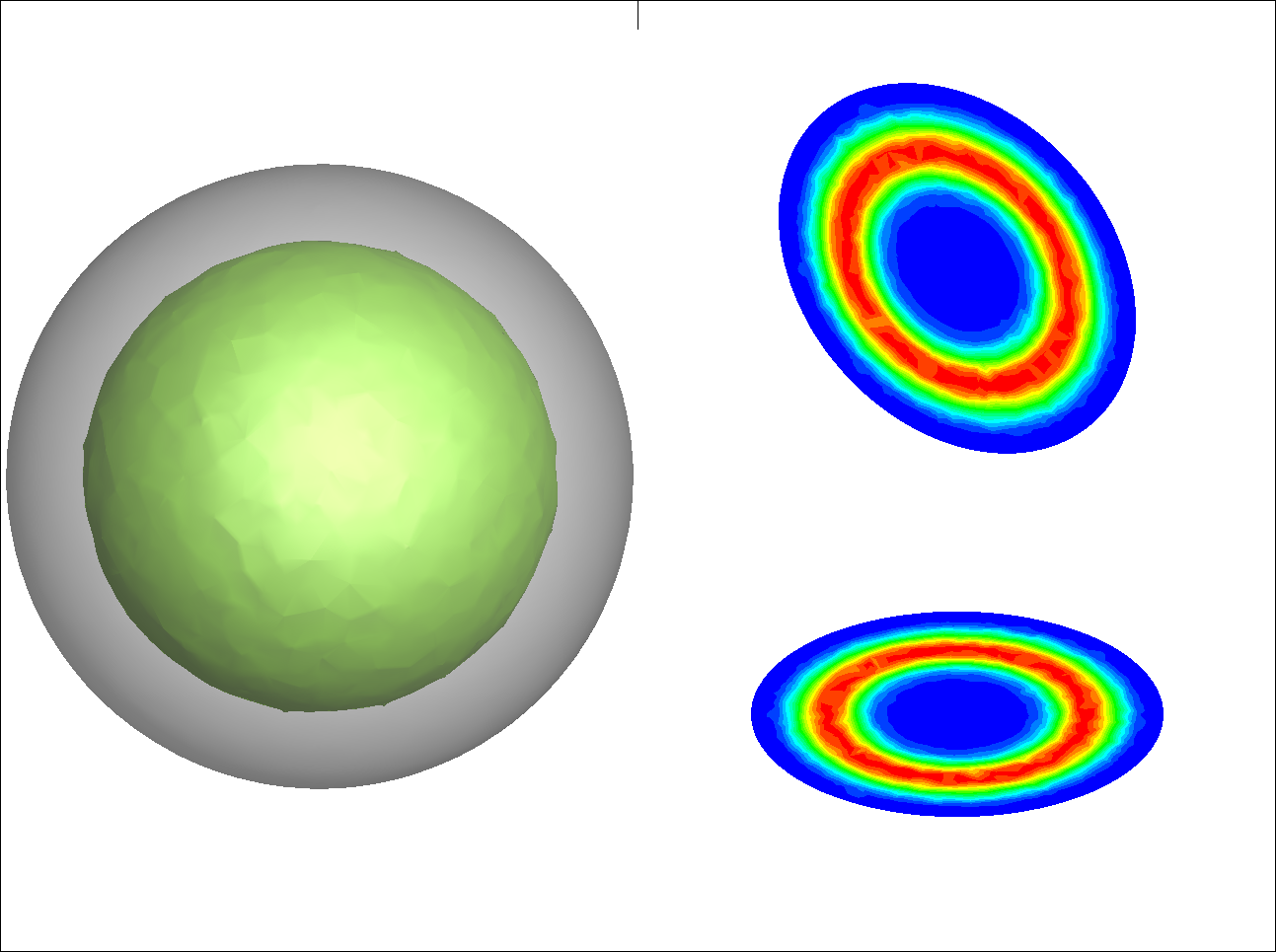} \hspace*{-0.9em}
\includegraphics[width=1.2in]{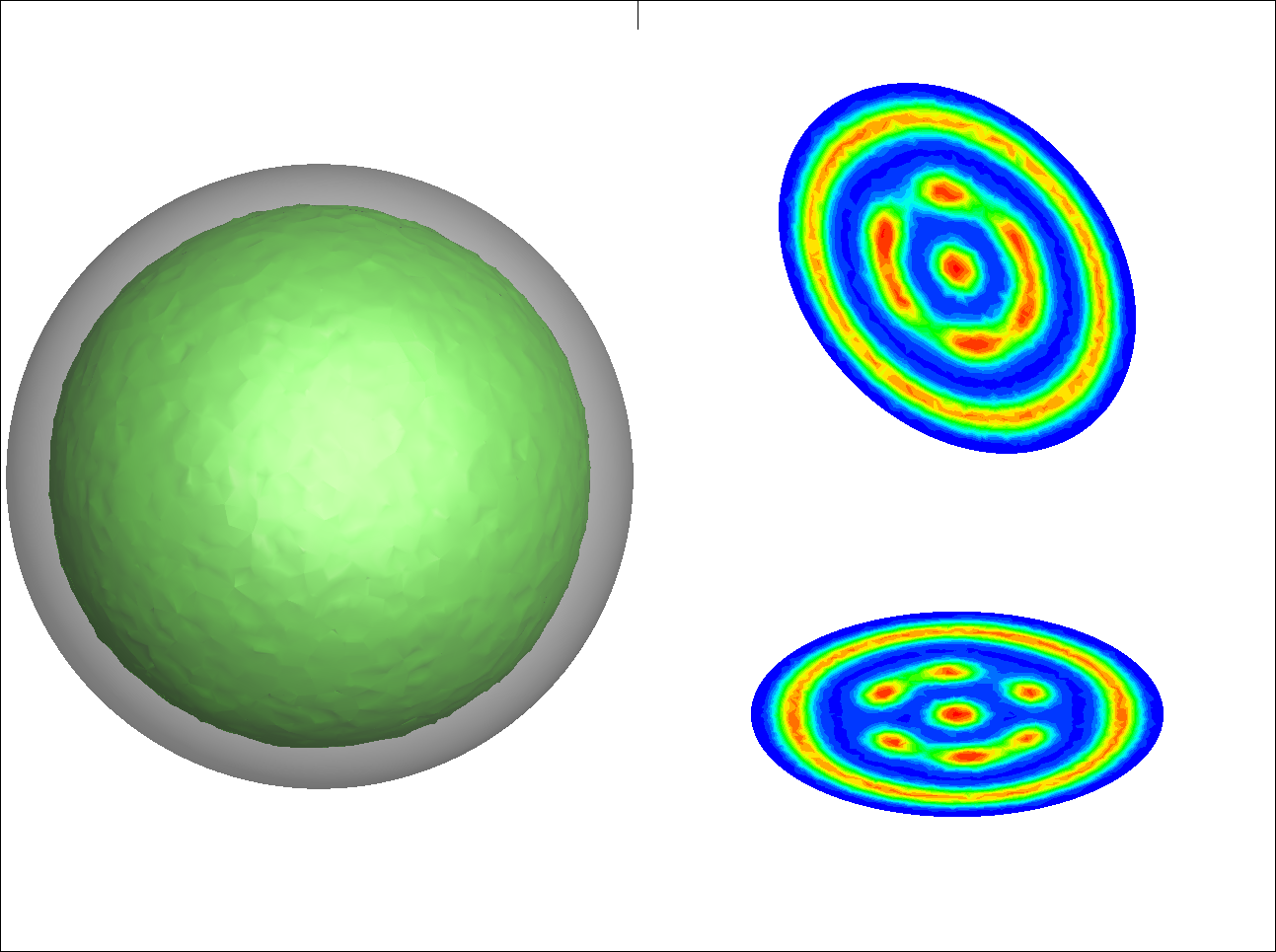}\\

\caption{\label{F:fig_15} Equilibrium microstructures at $f_{A}=30$ and $\chi N=18$ for varying geometries and confinement volume. From top to bottom, the confinement geometry changes from a tetrahedron to sphere based on curvature driven flow with conserved volume. From left to right the volume of the confinement geometry increases, corresponding to a tetrahedral volume of edge length, $L/4$, $L/2$, $L$ and $2L$ with $L=14.72R_{g}$. The corresponding confinement volumes for each geometry are $V_{1}$, $V_{2}$, $V_{3}$ and $V_{4}$ respectively. The legend at the top indicates the location of cross-section for the 2D slices shown in each microstructure. In the 3D view, the green is the isosurface where $\rho_A=0.5$. In the cross sections, blue corresponds to B block while red is the A block.
}
\end{center}
\end{figure*}

First, we consider the case when $f_{A} = 30$ and analyze the effects of varying curvature and volume. Figure~\ref{F:fig_15} shows the equilibrium microstructures obtained at $f_{A}=30$ for different confinement volumes and the five geometries of varying curvature described above. Moving from left to right in the figure, the tetrahedron edge lengths increase from $\frac{L}{4}$ to $2L$, with the corresponding volumes increasing from $V_{1}$ to $V_{4}$. Going from top to bottom, the confinement geometries change from a tetrahedron to a sphere. As noted above, for a confinement volume of $V_{3}$ the equilibrium microstructure is interpenetrating for the tetrahedron geometry. As the confinement geometry changes to a sphere, we see from the isocontours and slices that the $A$ component becomes more connected. In the bottom two geometries of column 3, the outer $B$ layer is fully separated from the inner $B$ layer leading to a discretely disconnected microstructure.
Increasing the volume of tetrahedron to $V_{4}$ (column 4), the outer $B$ component is again fully connected to the innermost region. Compared to the microstructure with volume $V_{3}$, we can see a more interesting structure with interconnected  rods for the structure of the $A$ component. As the geometry changes to a sphere, the outer layer of the $A$ component transitions to a concentric shell with the outermost $B$ component completely disconnected from the inner $B$ component. Inside, a series of concentric shells forms, although the inner A shells are not fully connected at this value of $f_{A}$.

Exploring the effect of smaller geometries, we look at column 2 where the confinement volume is $V_{2}$. For the tetrahedron, the equilibrium microstructure consists of only the outer $B$ component and an inner $A$ component that adopts a shape reflecting the confinement geometry. This same behaviour is seen in all the shapes. As the confinement volume of the tetrahedron is further reduced to $V_{1}$ (column 1), we find that the inner $A$ component is pushed away from the edges and corners where the field is strongest and towards the surface where the effect of the wall field is smaller. This creates an inner structure similar to that for confinement volume equal to $V_{2}$, but with more rounded edges and corners. This effect becomes less pronounced as the geometries tend to a sphere.

\begin{figure*}
\begin{center}
\includegraphics[width=1.2in]{tet_d4} \hspace*{-0.9em}
\includegraphics[width=1.2in]{tet_d2} \hspace*{-0.9em}
\includegraphics[width=1.2in]{tet_d1} \hspace*{-0.9em}
\includegraphics[width=1.2in]{tet_m2}\\ 
\vspace{-0.08em}
\includegraphics[width=1.2in]{tet_f30_d4} \hspace*{-0.9em}
\includegraphics[width=1.2in]{tet_f30_d2} \hspace*{-0.9em}
\includegraphics[width=1.2in]{tet_f30_d1} \hspace*{-0.9em}
\includegraphics[width=1.2in]{tet_f30_m2}\\
\vspace{-0.08em}
\includegraphics[width=1.2in]{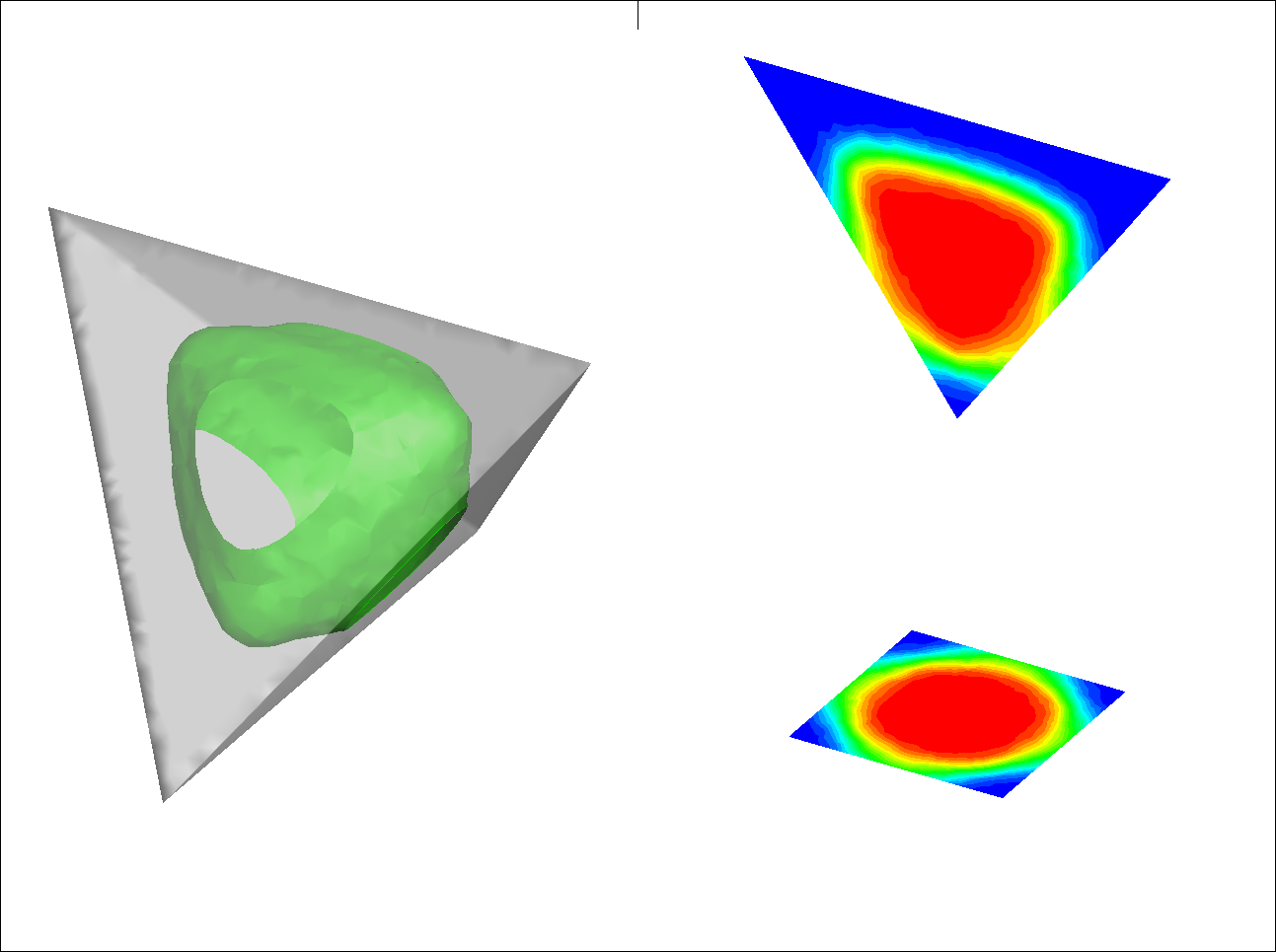} \hspace*{-0.9em}
\includegraphics[width=1.2in]{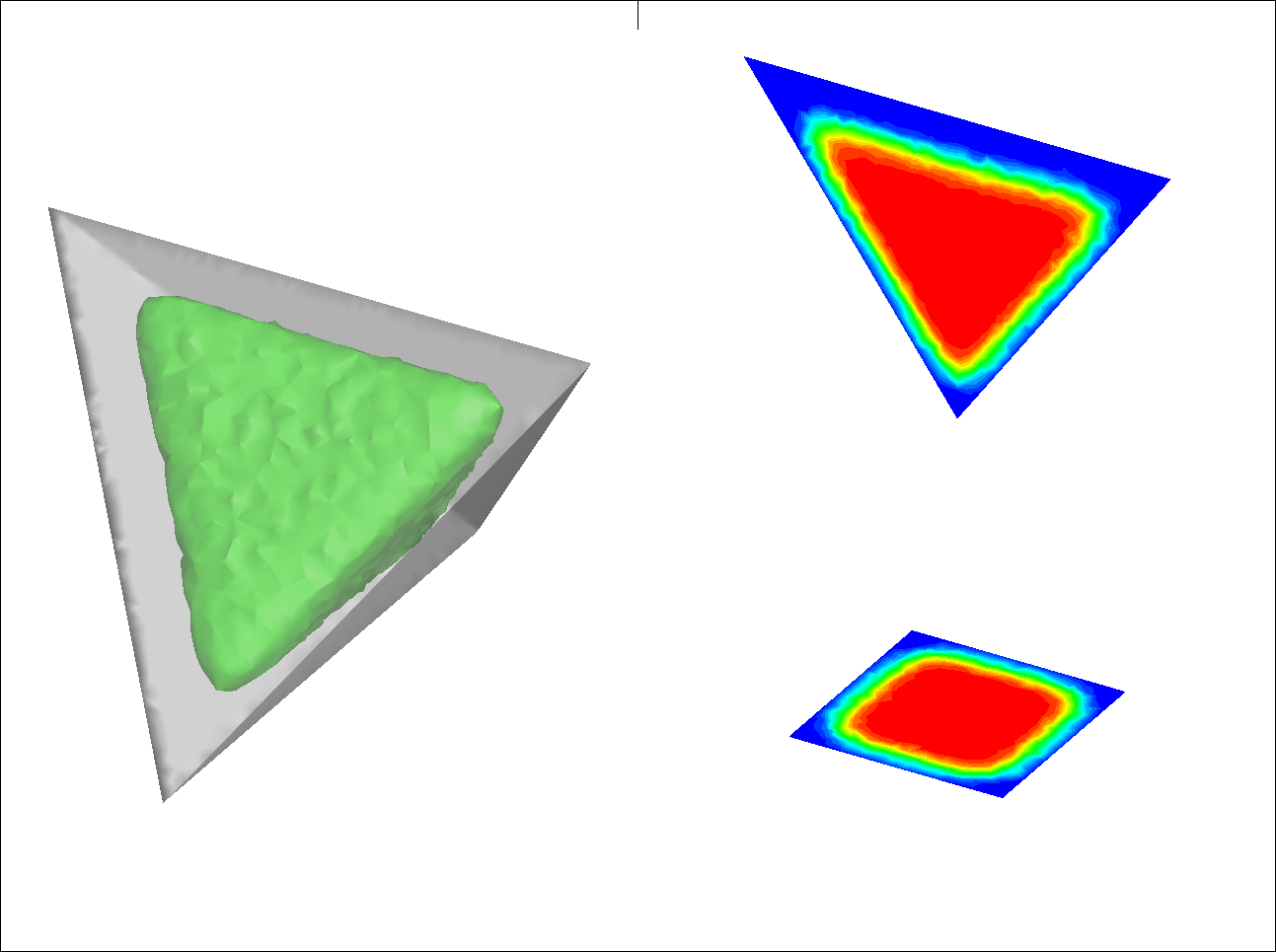} \hspace*{-0.9em}
\includegraphics[width=1.2in]{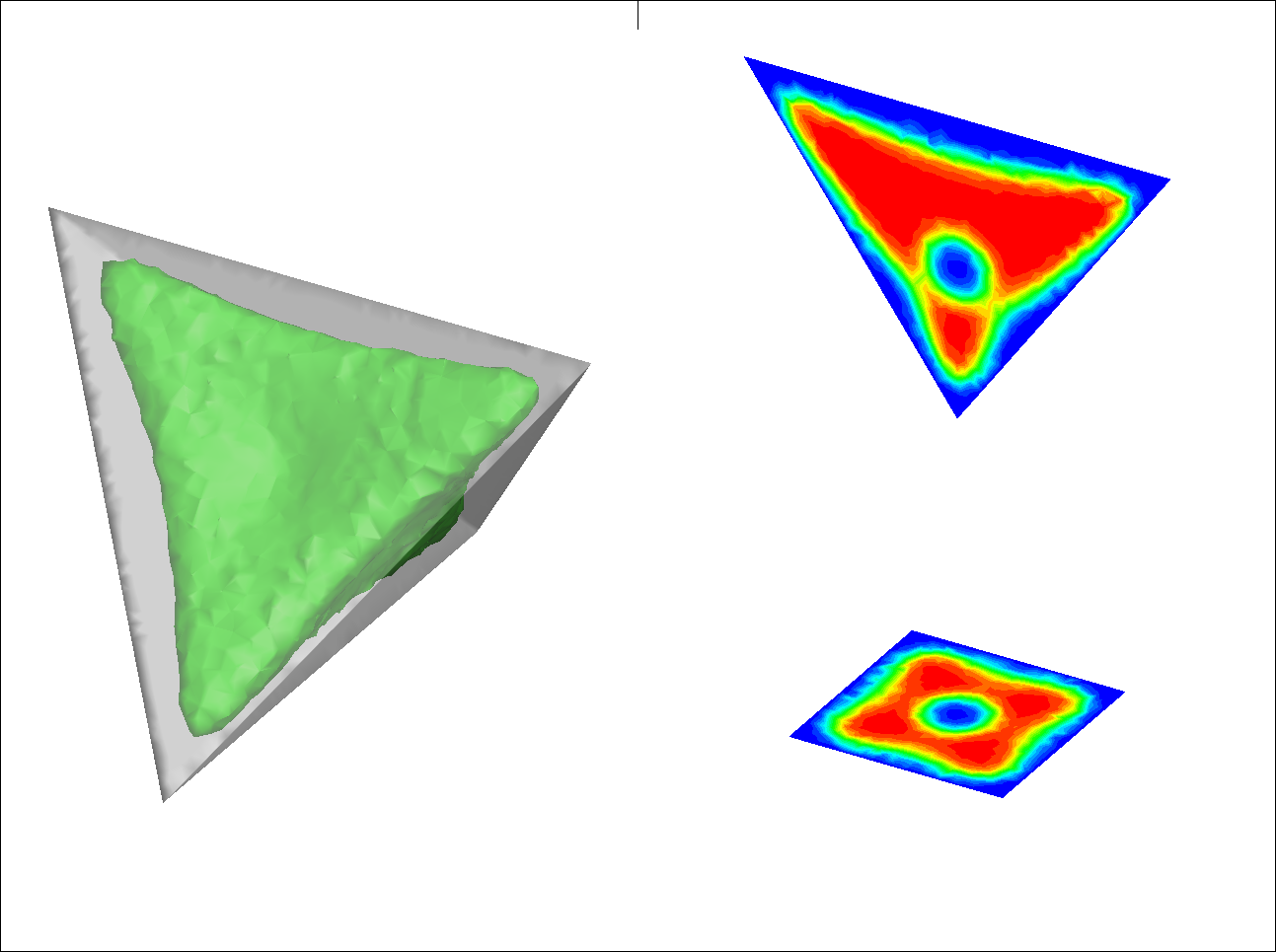} \hspace*{-0.9em}
\includegraphics[width=1.2in]{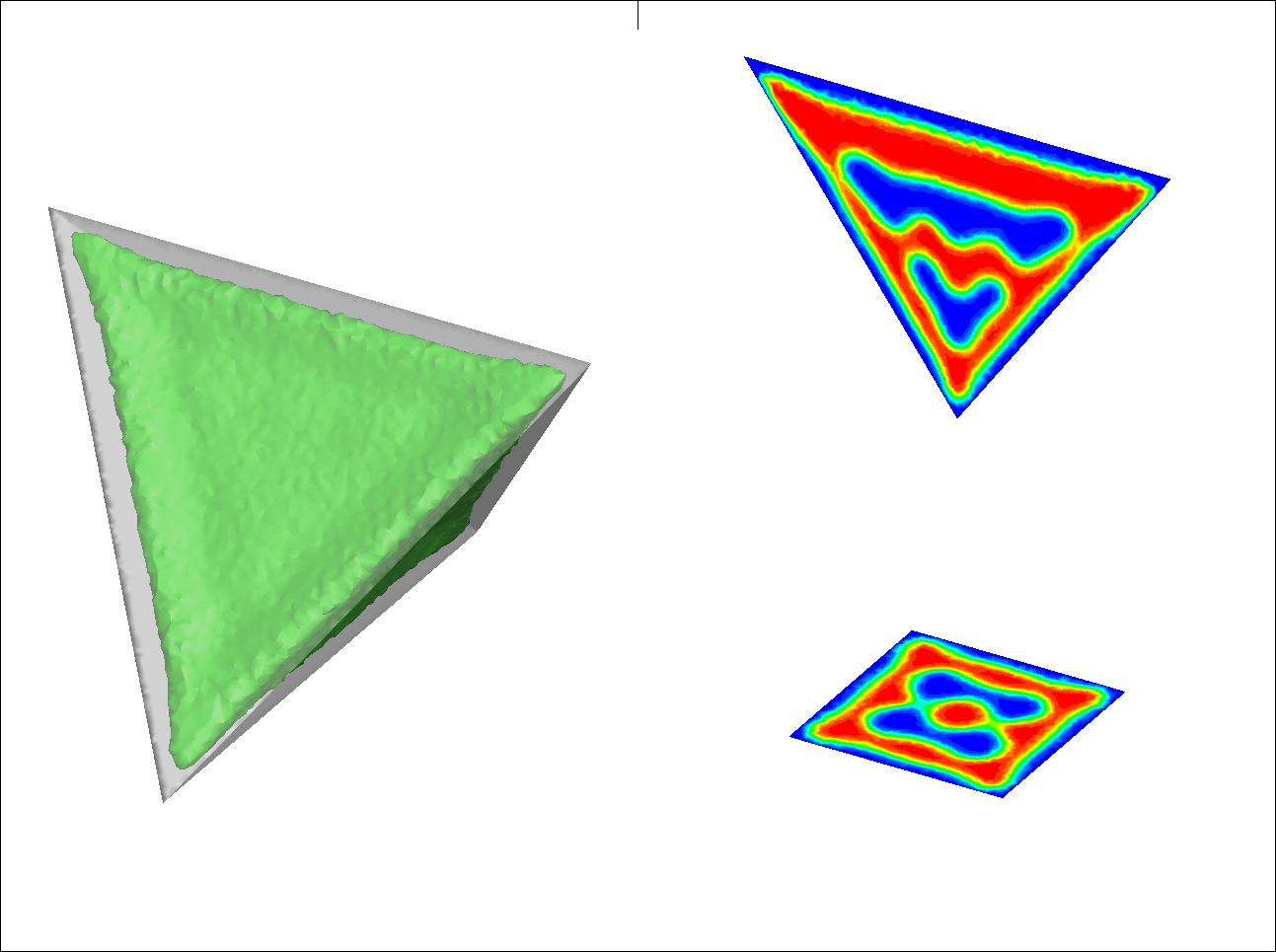}\\
\vspace{-0.08em}
\includegraphics[width=1.2in]{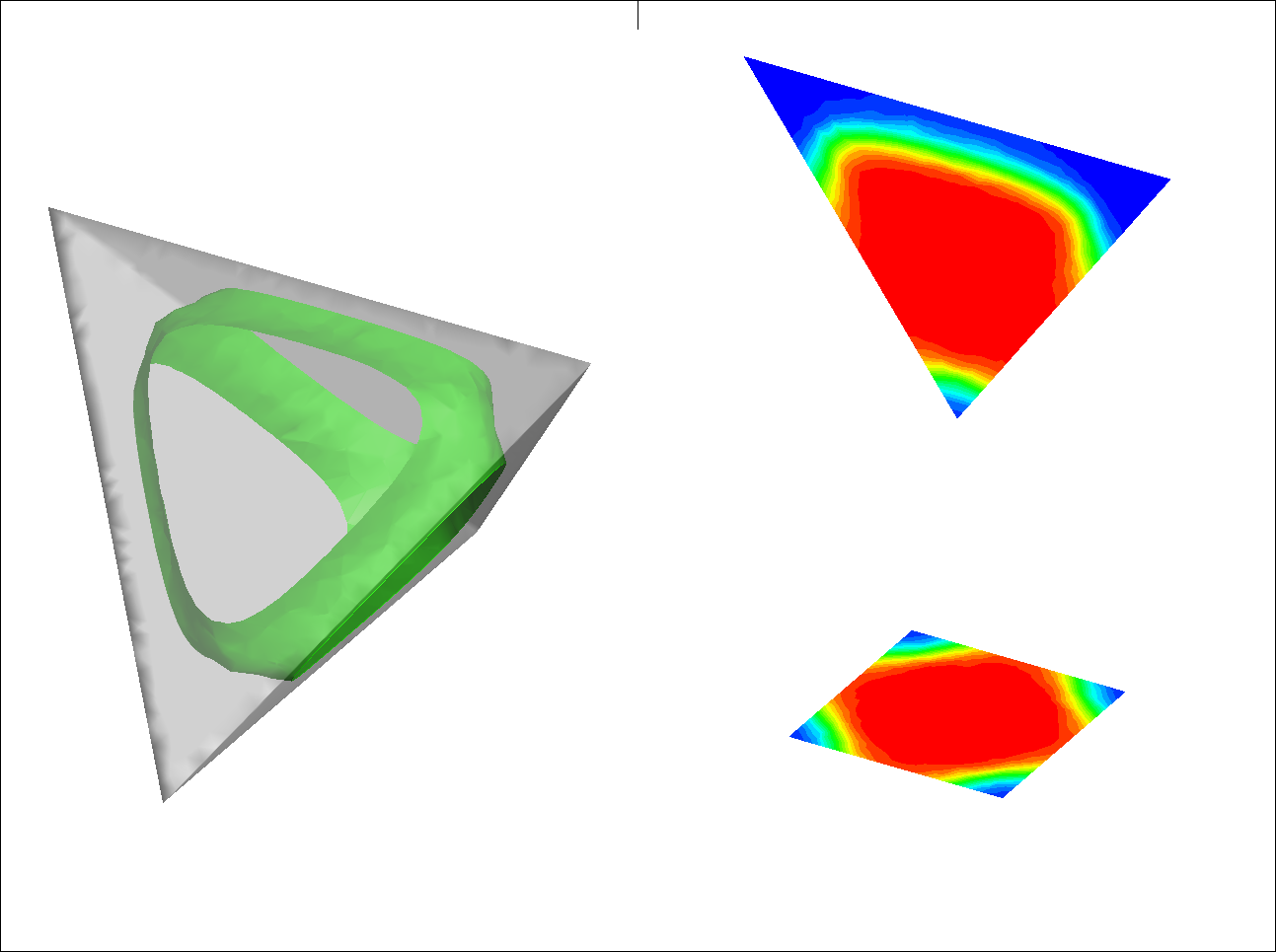} \hspace*{-0.9em}
\includegraphics[width=1.2in]{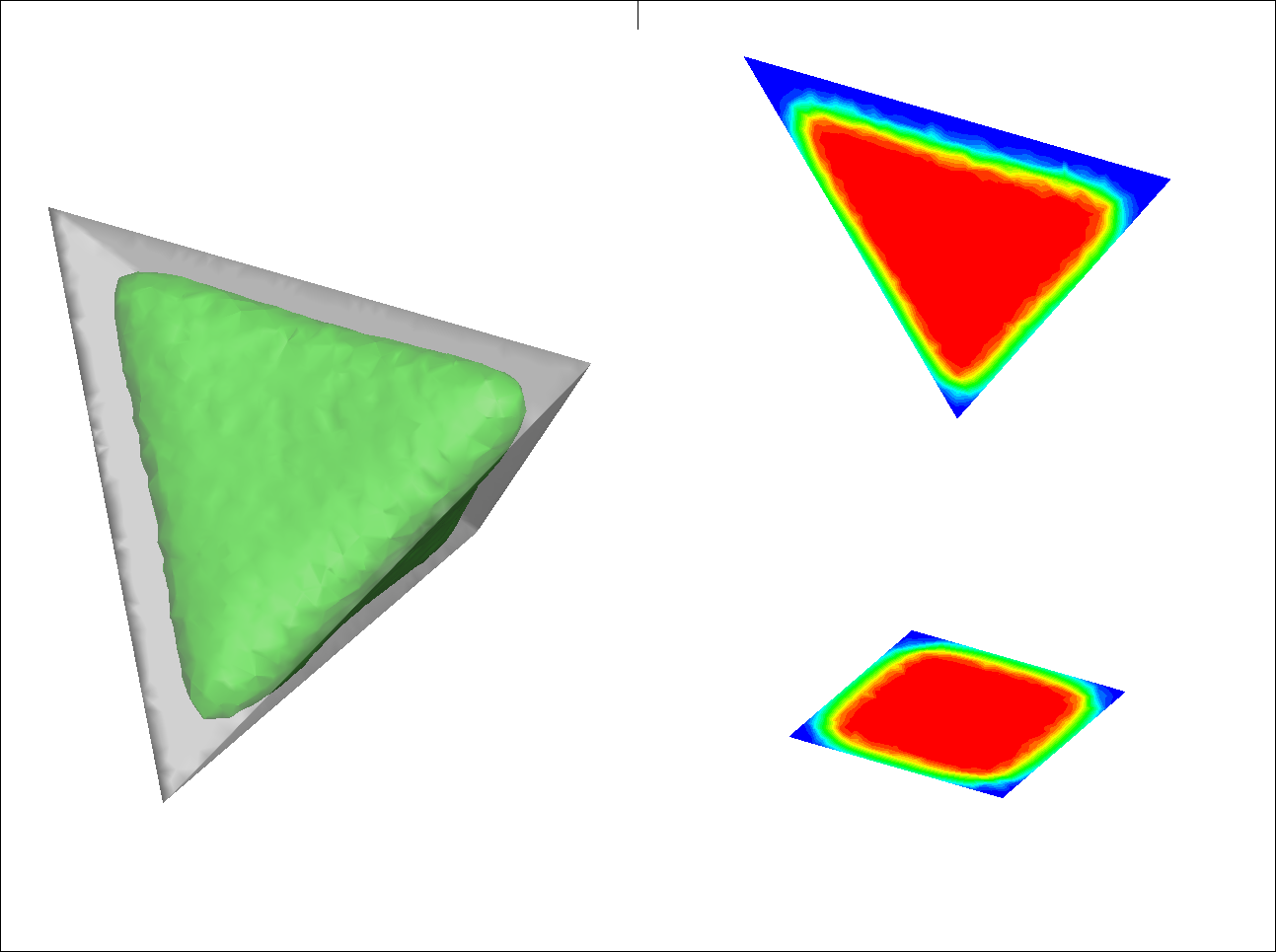} \hspace*{-0.9em}
\includegraphics[width=1.2in]{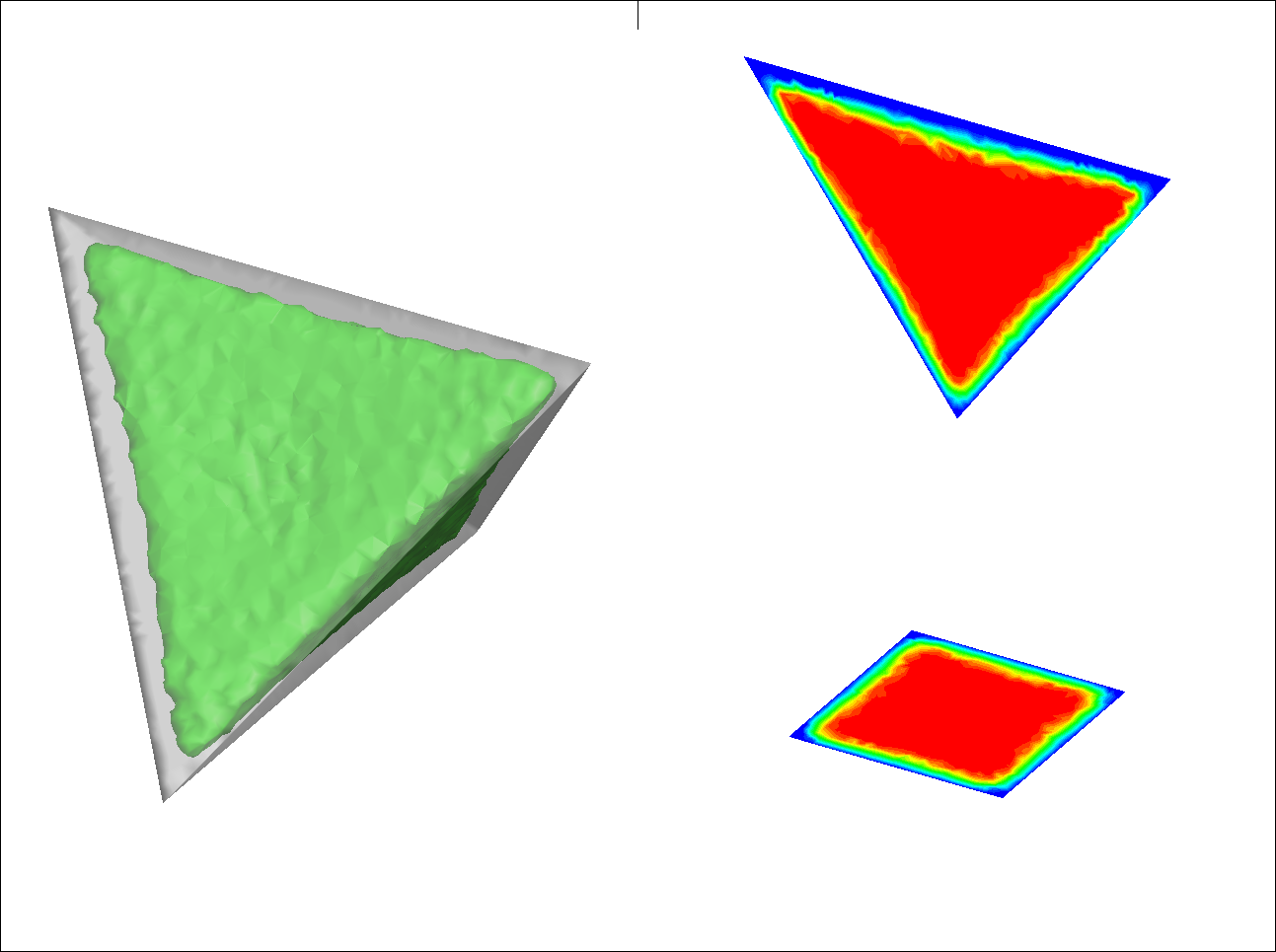} \hspace*{-0.9em}
\includegraphics[width=1.2in]{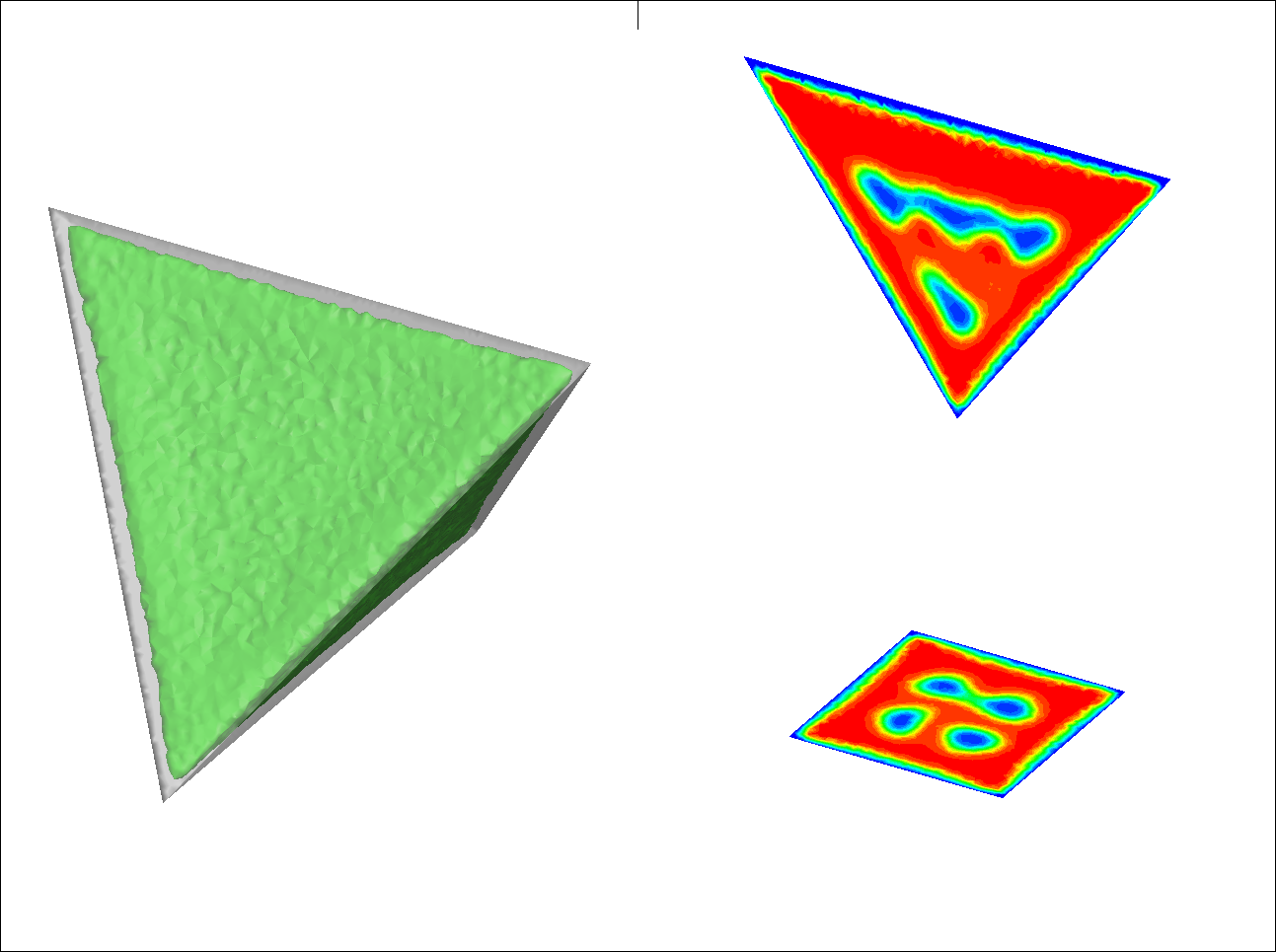}\\
\caption{\label{F:fig_16} Equilibrium microstructures at $f_{A}=30, 50, 70$ and $\chi N=18$ for a tetrahedral geometry with varying size of confinement volume. From top to bottom, the volume fraction along each row is $f_{A}=30$, $f_{A}=50$, $f_{A}=70$ respectively. From left to right the confinement
volume is $V_{1}$, $V_{2}$, $V_{3}$, and $V_{4}$, corresponding to tetrahedral edge lengths, $L/4$, $L/2$, $L$, and $2L$ respectively, where $L=14.72R_{g}$
}
\end{center}
\end{figure*}

\subsection{Variation of volume fraction and confinement volumes with fixed curvature}

\begin{figure*}
\begin{center}
\includegraphics[width=1.2in]{tet_d4} \hspace*{-0.9em}
\includegraphics[width=1.2in]{tet_d2} \hspace*{-0.9em}
\includegraphics[width=1.2in]{tet_d1} \hspace*{-0.9em}
\includegraphics[width=1.2in]{tet_m2}\\
\vspace{-0.08em}
\includegraphics[width=1.2in]{curv6_f30_d4} \hspace*{-0.9em}
\includegraphics[width=1.2in]{curv6_f30_d2} \hspace*{-0.9em}
\includegraphics[width=1.2in]{curv6_f30_d1} \hspace*{-0.9em}
\includegraphics[width=1.2in]{curv6_f30_m2}\\
\vspace{-0.08em}
\includegraphics[width=1.2in]{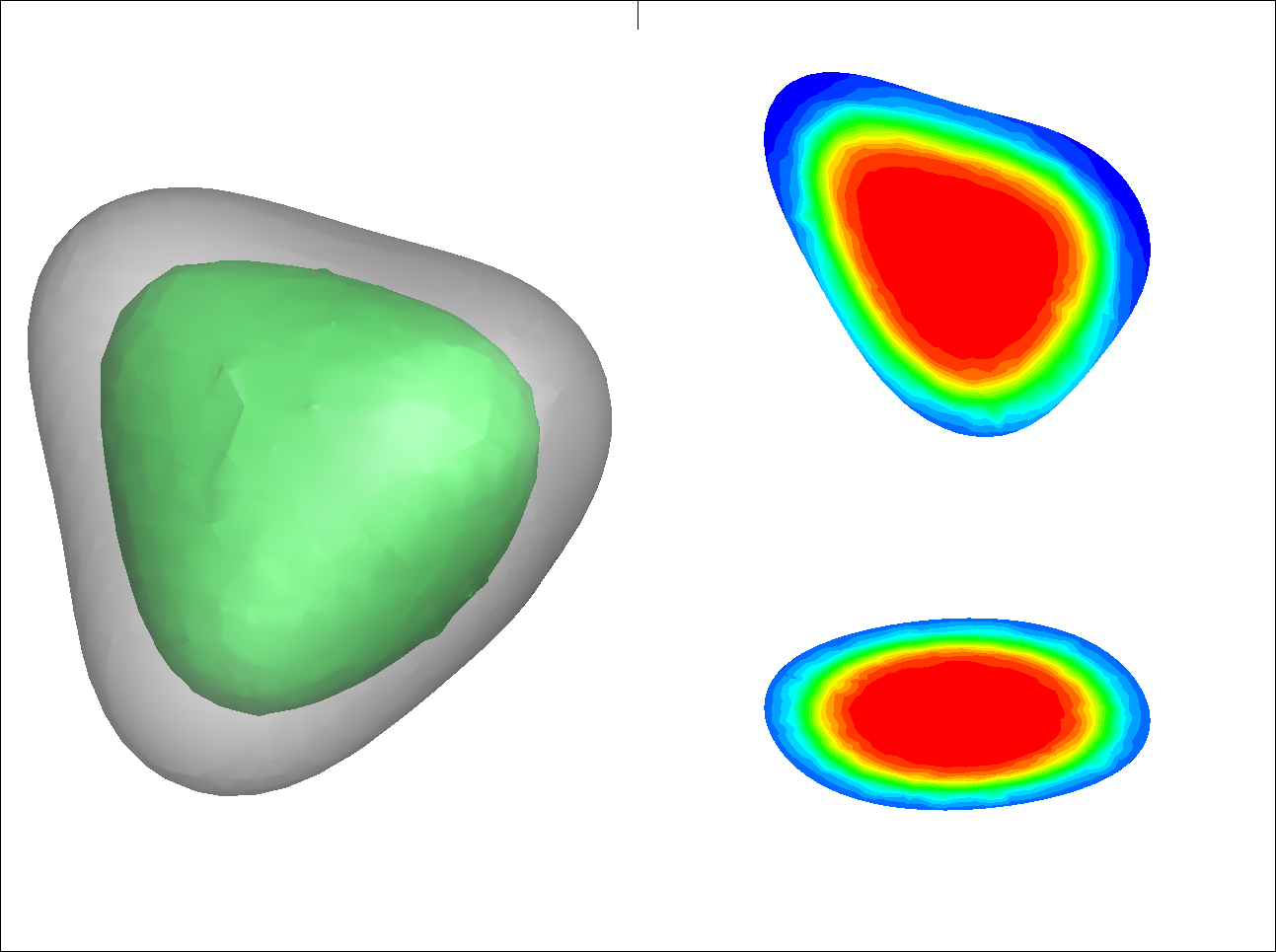} \hspace*{-0.9em}
\includegraphics[width=1.2in]{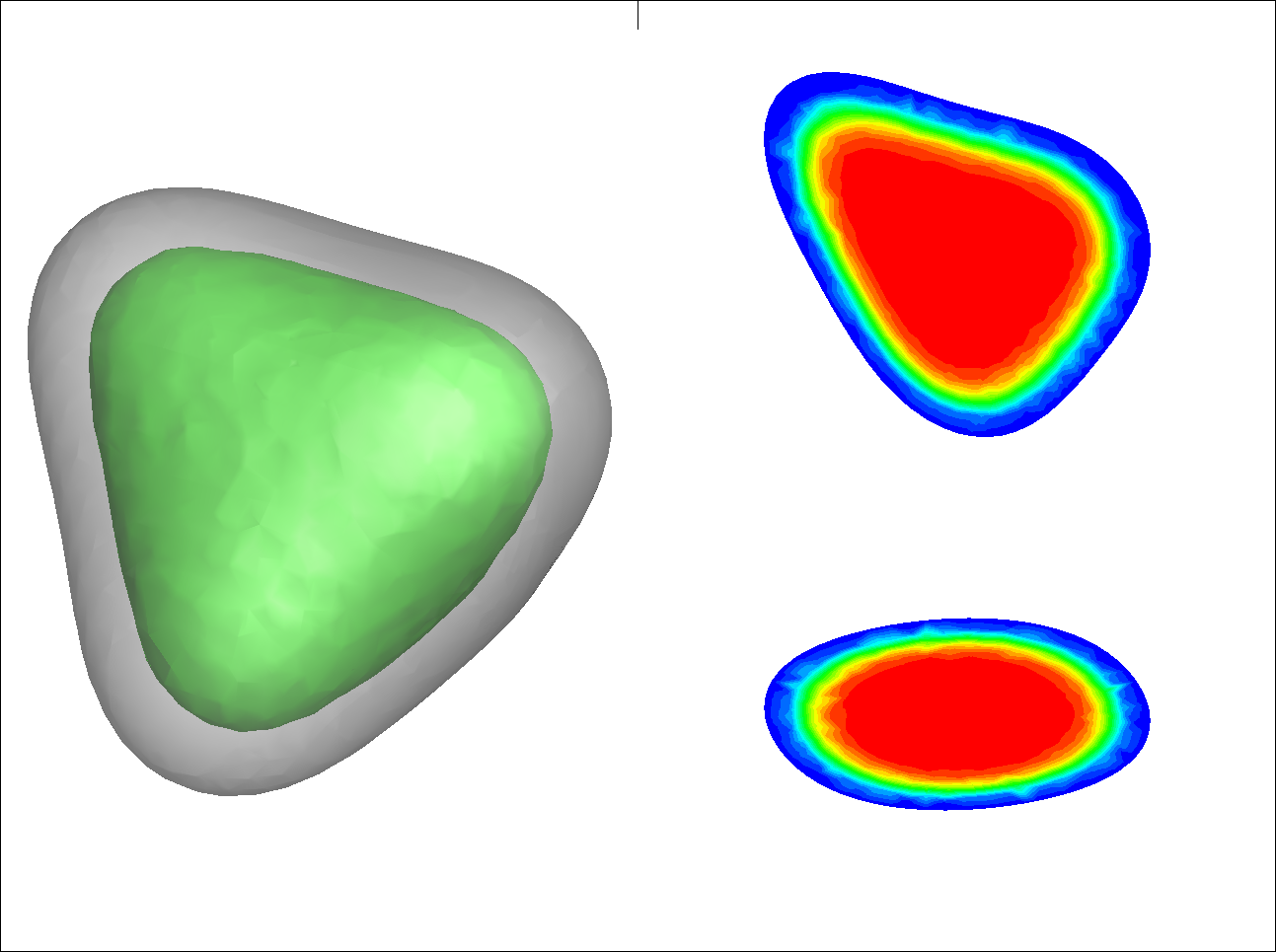} \hspace*{-0.9em}
\includegraphics[width=1.2in]{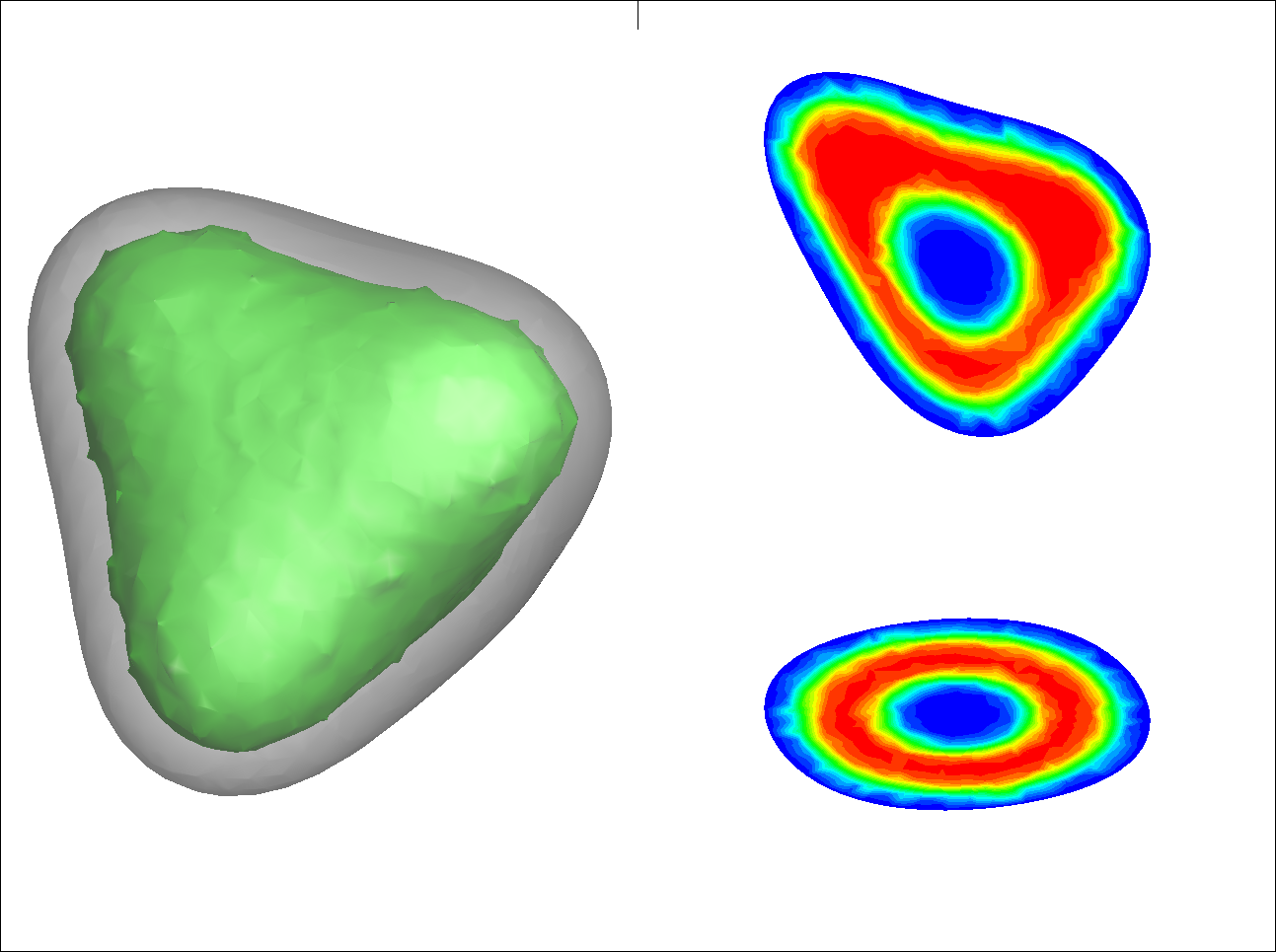} \hspace*{-0.9em}
\includegraphics[width=1.2in]{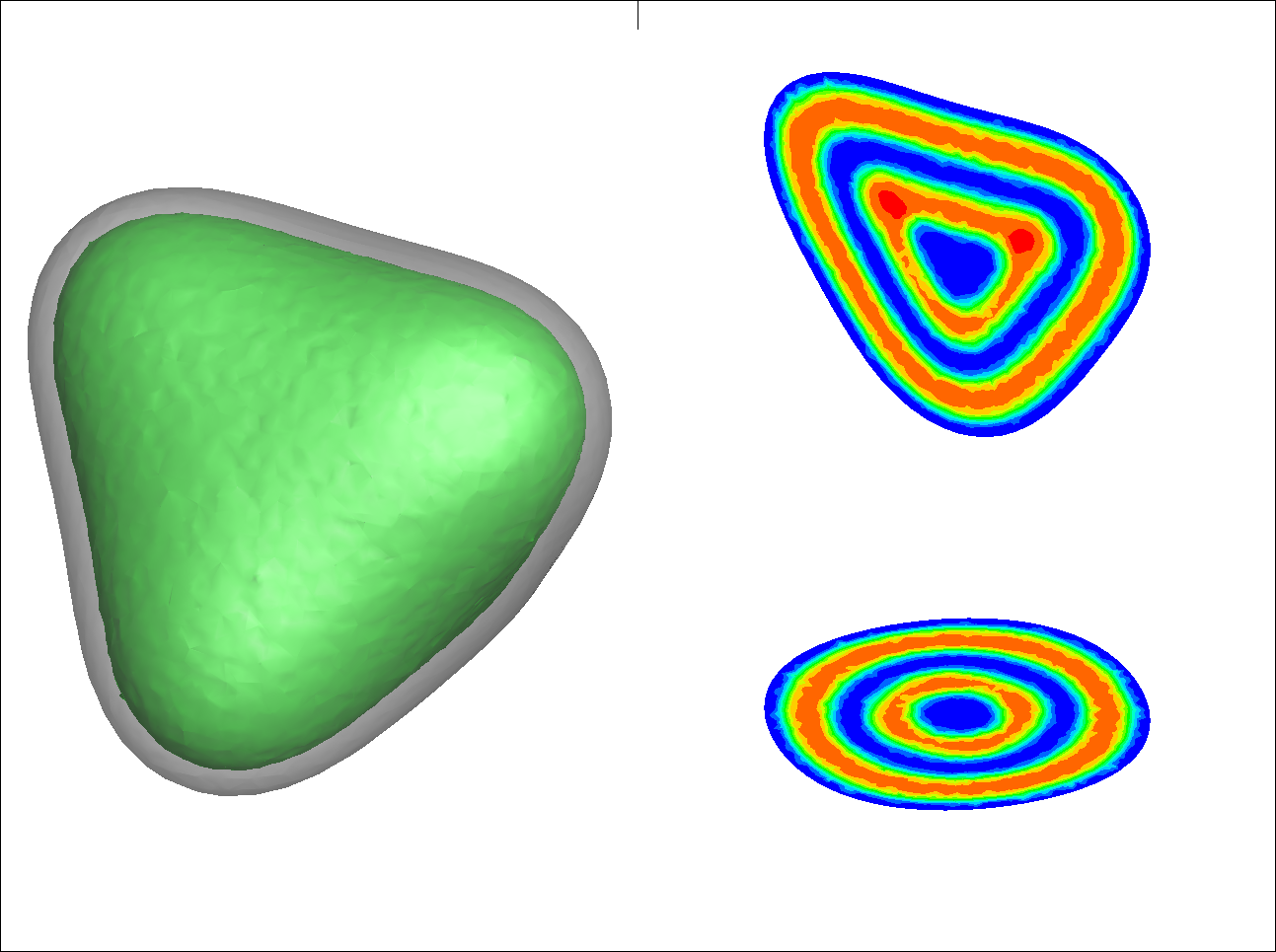}\\
\vspace{-0.08em}
\includegraphics[width=1.2in]{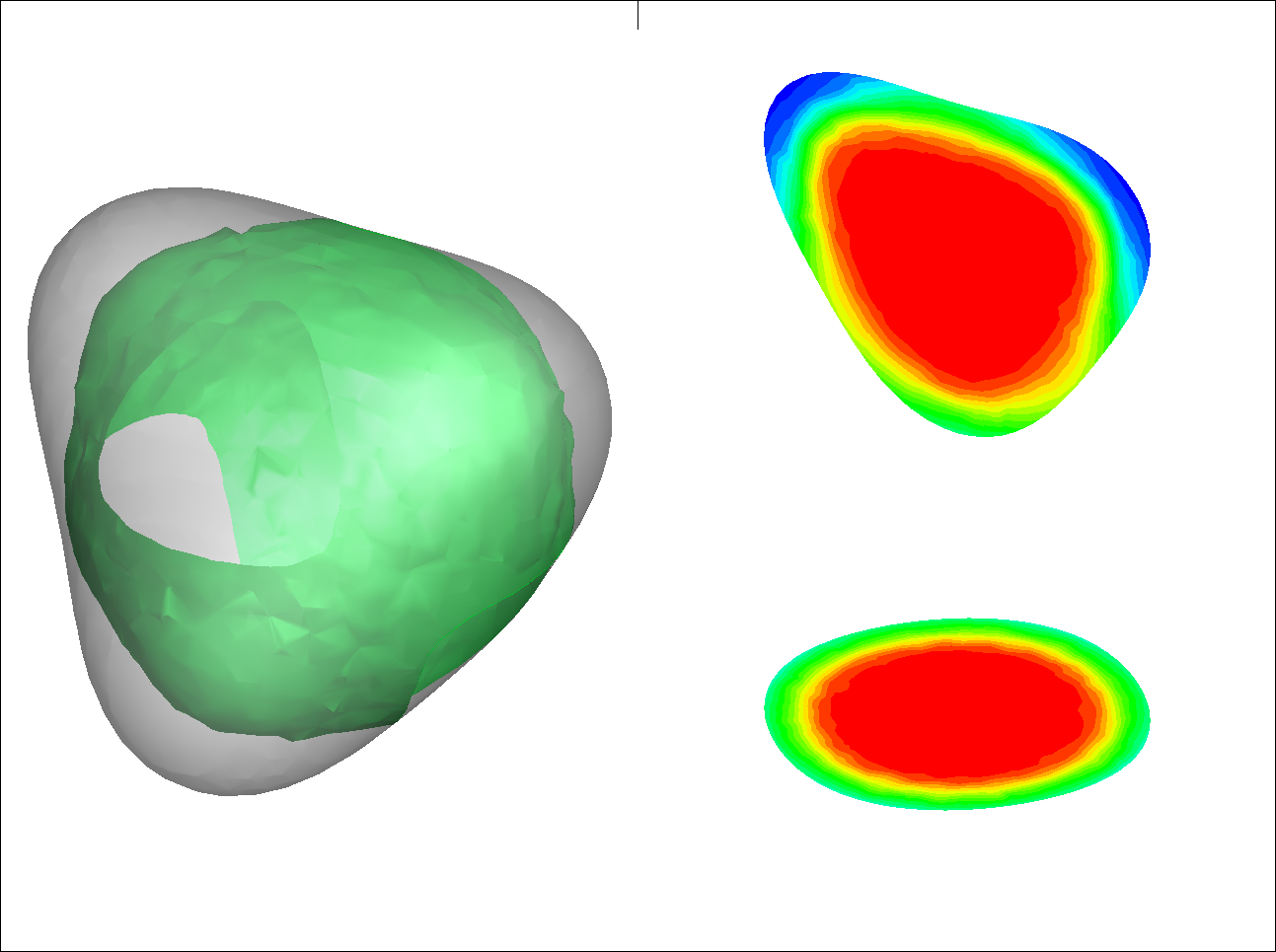} \hspace*{-0.9em}
\includegraphics[width=1.2in]{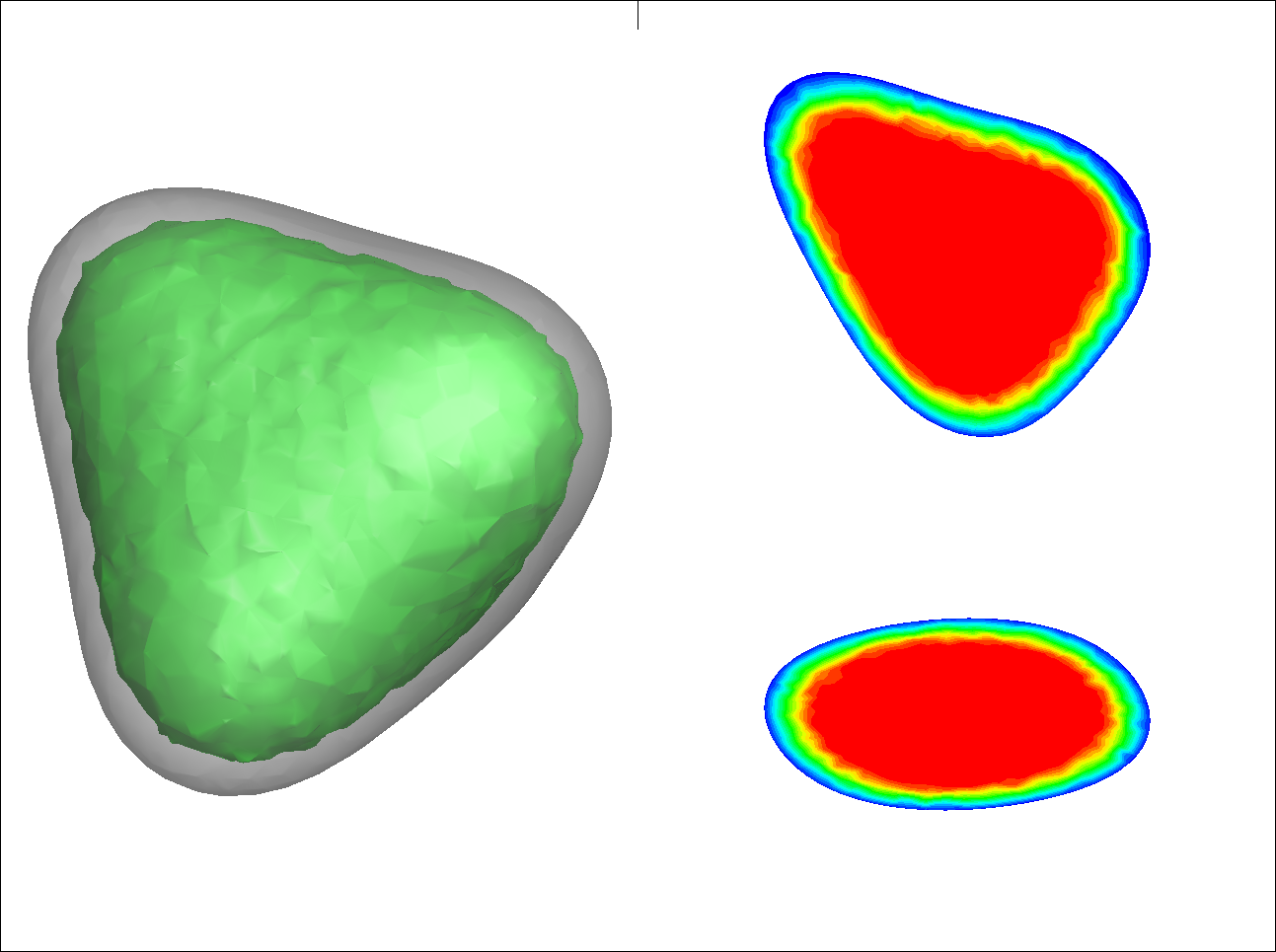} \hspace*{-0.9em}
\includegraphics[width=1.2in]{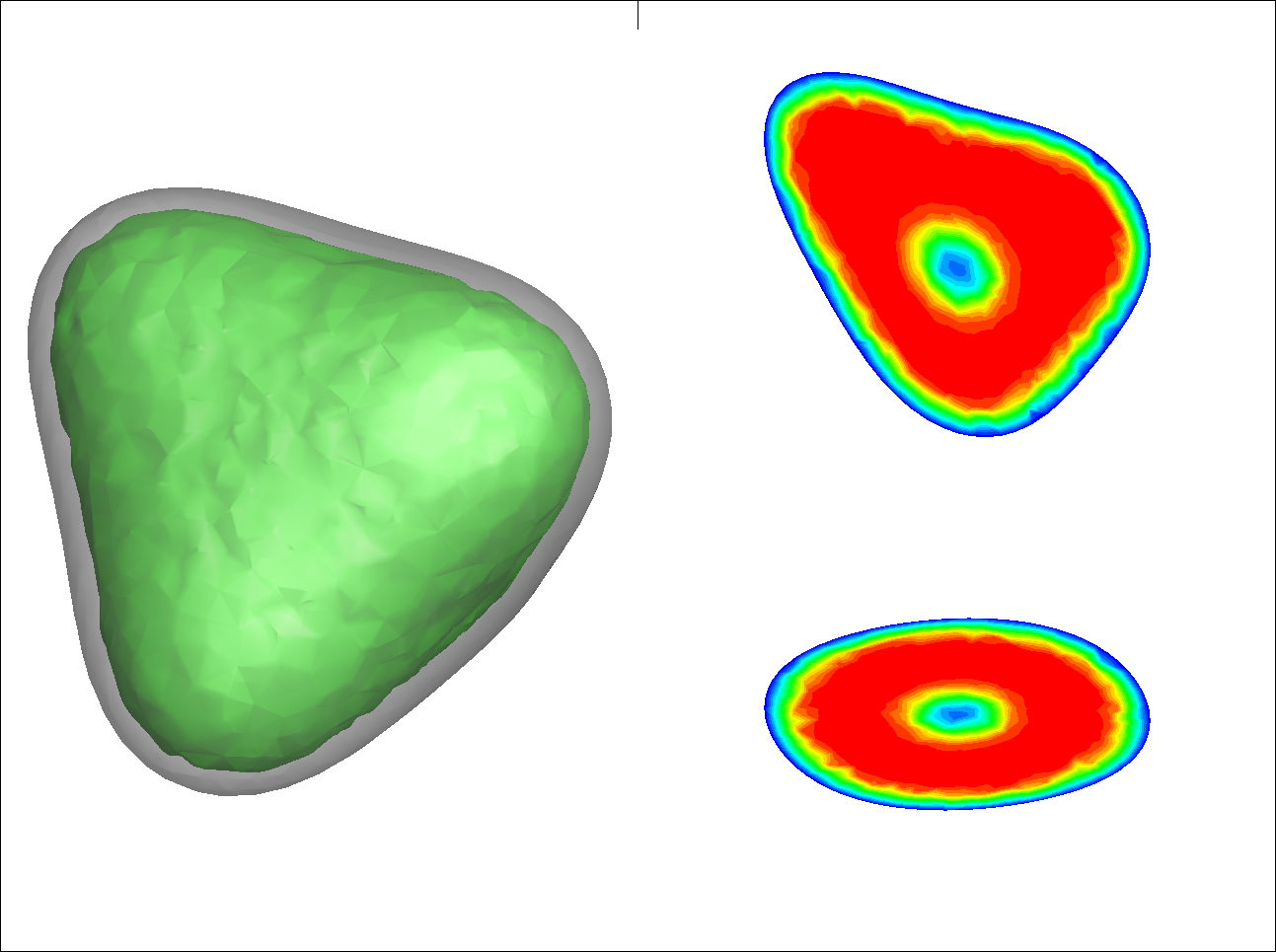} \hspace*{-0.9em}
\includegraphics[width=1.2in]{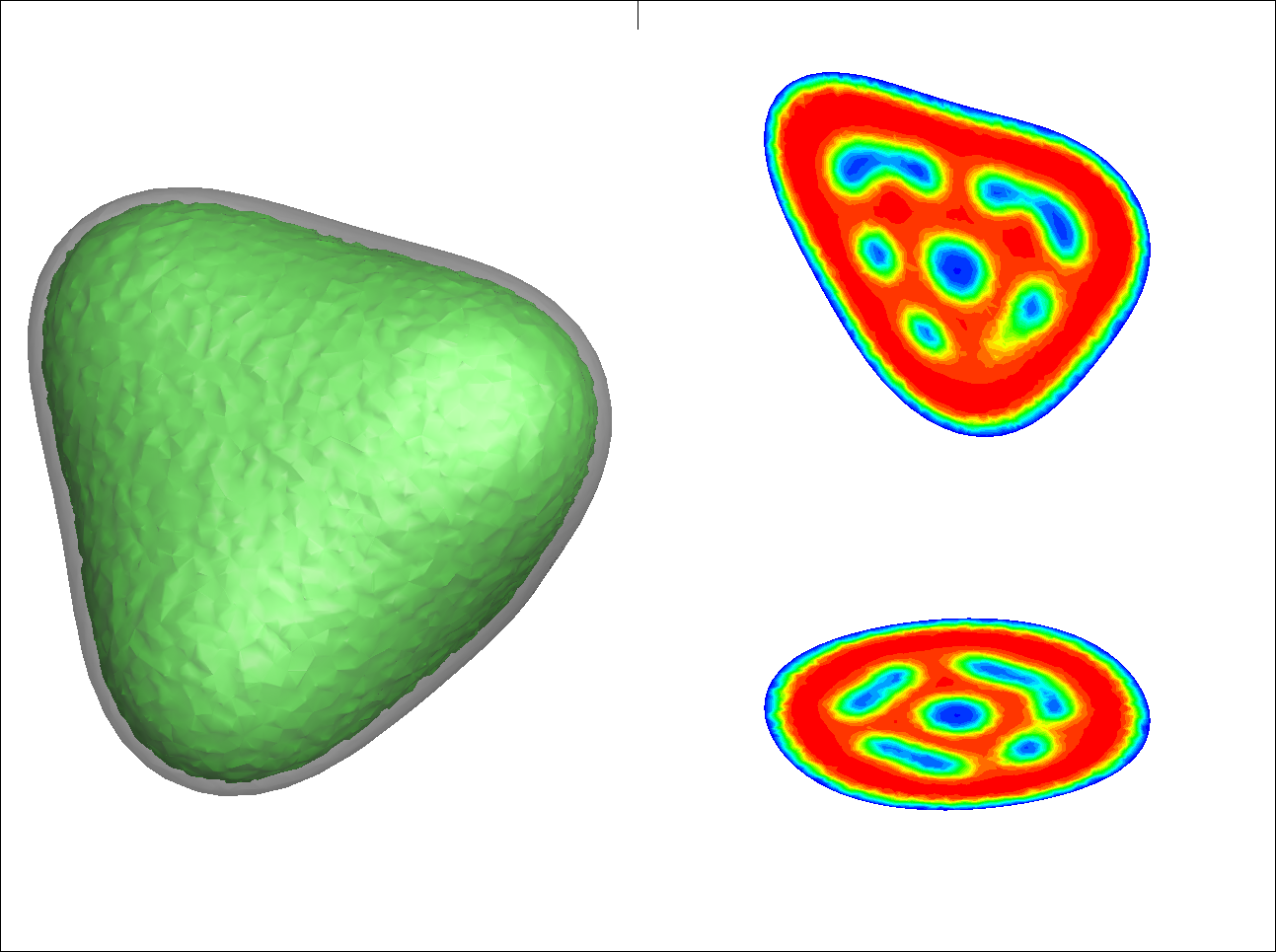}\\
\caption{\label{F:fig_17} Equilibrium microstructures at $f_{A}=30, 50, 70$ and $\chi N=18$ for an intermediate geometry with varying size of confinement volume. From top to bottom, the volume fraction along each row is $f_{A}=30$, $f_{A}=50$, $f_{A}=70$ respectively. From left to right the volume of the confinement geometry increases. The confinement volumes are $V_{1}$, $V_{2}$, $V_{3}$, and $V_{4}$, corresponding to tetrahedral edge lengths, $L/4$, $L/2$, $L$, and $2L$ respectively, where $L=14.72R_{g}$.
}
\end{center}
\end{figure*}

We also consider the effect of changing volume fractions, $f_{A}$, on the equilibrium microstructures. We present the results for the tetrahedron and one intermediate geometry while changing $f_{A}$ and the volume. Figure~\ref{F:fig_16} shows the microstructures for tetrahedrons with $f_{A} = 30, 50, 70$. Moving from left to right, the edge length increasing from $L/4$ to $2L$, with the corresponding volume increasing from $V_{1}$ to $V_{4}$. Previous results from Figure~\ref{F:fig_15} correspond to the top row. For the baseline edge length L with volume $V_{3}$ (column 3), increasing $f_{A}$ from $30$ to $50$ leads to a transition from an interpenetrating to non-interpenetrating structure. Further, in the inner most region at $f_{A}=50$, we find a layer of $B$ component which is similar to spherical in shape. For $f_{A}=70$ the inner phase is fully composed of the $A$ component with a thin outer layer of $B$. Increasing the size to $2 L$ (column 4), at $f_{A}=50$, we find a thin outer layer of $B$ component which is attractive to the wall. Inside is an interpenetrating structure very similar to the $B$ structure for $f_{A}=30$ at volume $V_{3}$. At $f_{A}=70$ the thickness of the outer $B$ layer and inner $B$ structure are reduced. This corresponds to the connectivity of the internal $B$ component beginning to break up.  Going to smaller volumes, we see that a reduction in the volume to $V_{2}$ (column 2) gives a similar structure at $f_{A}=30, 50$ and $70$. However, a further reduction in the volume to $V_{1}$ leads to changes in the microstructures due to the effects of wall field becoming significant. At $f_{A}=70$, we see that the $A$ component is reaching the surface due to larger volume fractions at the regions where wall field is smaller. This effect is seen more clearly in the structures corresponding to volume, $V_{1}$. We can clearly see that the isosurface at $\rho _{A}=0.5$ shown in the figure is not a closed surface for $f_{A} =50, 70$. 
Instead, the density of the $A$ component in the open region is much higher ($\rho _{A} > 0.9$). This can be seen more clearly from the contour slices, where the red color represents a higher density of the $A$ component.
The outer region near the vertices of the tetrahedrons are still dominated by the $B$ component mainly due to the stronger wall fields which are attractive to the $B$ polymer segment. As we change the confinement geometry and move to a more uniform curvature, this effect will be reduced. This can be seen in the left most column of Figure~\ref{F:fig_17}, where we show the equilibrium microstructures for an intermediate geometry (corresponding to the $3^{rd}$ row in Figure~\ref{F:fig_15}). At $f_{A} = 50$, we see a fully connected structure for the isosurface at $\rho _{A}=0.5$ (closed surface for the isosurface at $\rho _{A} = 0.5$) 
and at $f_{A}=70$, we see the the $A$ structure penetrates the outer $B$ shell near the center of the triangular face where wall field is smaller than the edges. Increasing the volume to $V_{2}$ (column 2) for this geometry leads to a solid inner $A$ layer and an outer $B$ shell at all volume fractions. At volume $V_{3}$ (column 3) for $f_{A}=50,70$, we get a non-interpenetrating concentric shell microstructure. Considering the volume, $V_{4}$ at $f_{A}=50$, we see the formation of concentric layers of $A$ and $B$ components. Increasing the volume fraction to $f_{A}=70$ breaks the concentric layer of $B$ component in the middle.

\section{Conclusions} \label{S:concl}
We use a finite-element based, parallel, self-consistent field theoretic (SCFT) framework to generate equilibrium microstructures under 3D confinement. By varying the confinement geometry from a tetrahedron to sphere while maintaining a constant volume, we explored the interplay between curvature and confinement on the equilibrium microstructures. We considered variations in volume fraction of the diblock copolymer as well as the variation in volume of the confinement geometry. For a given volume, $V_{3}$ corresponding to a tetrahedron with edge length $L=14.72 R_{g}$, we find that the microstructure transitions from an interpenetrating structure to concentric spherical shell as the geometry transitions from tetrahedron to sphere at $f_{A}=30$. This effect is also seen at lower values of volume fraction, $\sim f_{A}=30$. Increasing the volume fraction gradually leads to the A component being confined to the interior of the volume even for the tetrahedral geometry. Although most structures had a solid outer layer of $B$ at the edges due to the attractive wall field, we find that decreasing the edge length by a factor of $4$ leads to the $A$ component from the interior of the tetrahedral penetrating the outer $B$ layer to touch the edge in the center of the faces. This is presumably energetically preferable to expanding to fill the areas of high field values near the edges and corners of the non-spherical geometries. As the curvature changes to a sphere, this effect is reduced and the A component is again confined to the interior of spherical geometry even at large volume fractions due to uniform wall field across the surface. We also explored the effect of increasing the size by considering the tetrahedral volume,$V_{4}$ with size $2L$. The structure of the A component is more complex with inter-connected rods at $f_{A}=30$. As the geometry changes to a sphere with the confinement volume being constant at $V_{4}$, we find that the equilibrium microstructures transform to a multi-layered concentric shells of the A and B components.    

In general, we find that at smaller confinement volumes, the diblock copolymers under confinement microphase separate into only two distinct components at any geometry and volume fraction with the outer region being composed of the polymer component attractive to the wall. We observe more interesting structures and edge/curvature effects in larger confinement volumes. In our study, this volume would correspond to a tetrahedron edge length of $L=14.72 R_{g}$ (confinement volume, $V_{3}$) and higher. For geometries with sharp edges like a tetrahedron, we observe microstructures with interpenetrating networks, especially at smaller volume fractions. We speculate that realization of these structures will have interesting applications. For instance, the networked structures might find use in catalysis applications where high surface-to-volume ratio are needed \citep{doi:10.1021/nl0495256}. As the geometry changes from tetrahedral to sphere, there is a transition of the inter-penetrating structure to multi-layered concentric phase-separated microstructures. These could have various biological applications including nano capsules. We envision that this study helps to provide insight into the kind of microstructures that can arise based on the confinement geometry involved and help to guide future experimental studies. Further computational exploration with other starting geometries might be helpful as the possibility of more complicated structures can offer new functionalities.

\bibliographystyle{elsarticle-num}
\bibliography{draft}

\appendix

\section{Energy convergence} \label{S:energyconv}

\begin{figure}
\begin{center}
\includegraphics[width=3.2in]{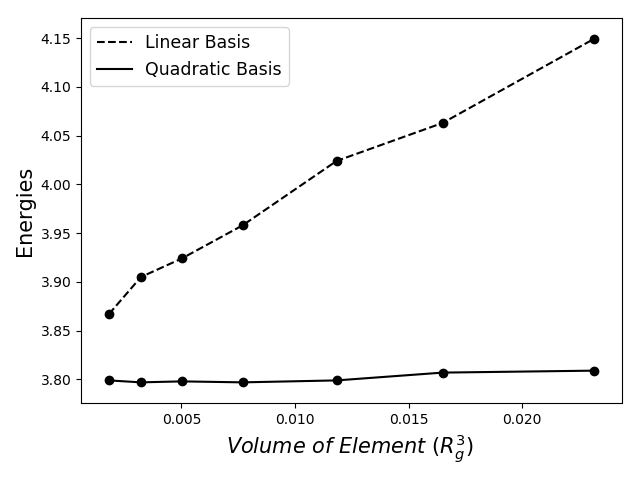}
\caption{\label{F:fig_energyconv} Energy convergence with element size.
}
\end{center}
\end{figure}

In principle, the SCFT calculation leads to lowest energy microstructure for given system parameters and initial conditions. However, this can be influenced by the resolution of the finite element mesh and choice of finite element basis. We choose mesh resolution such that the system energy is invariant as we further refine the mesh resolution. We perform a convergence analysis to obtain the minimum mesh resolution and the element basis required as well as the residual threshold needed for convergence. We performed SCFT calculations in a tetrahedral geometry with edge length $14.72 R_{g}$ at $f_{A} =40$ and $\chi N = 18$ for various mesh resolutions using elements with both linear and quadratic basis functions.  Simulations were allowed to converge such that the residual is below $10^{-3}$. The final energy for structures on meshes with varying element size is shown in Figure~\ref{F:fig_energyconv}. For linear elements, energy decreases noticeably as elements get smaller and we do not see a convergence in energies even at the highest resolution of $\sim$ $0.002 R_{g}^{3}$ volume per element. However, for elements with quadratic basis functions, we can clearly see that the energies have converged even at the coarsest resolution of $0.02 R_{g}^{3}$ volume per element. Based on these results, the equilibrium microstructures shown in the paper are created using meshes of elements with quadratic basis functions and average element sizes smaller than $0.02 R_{g}^{3}$ Additionally, we also find that for all these mesh resolutions, when the residual falls below $10^{-2}$ the variation in energy is negligible. So, we can conclude that SCFT calculations have converged once the residual is below $10^{-2}$ and fix a threshold value of $10^{-2}$ in the residual for the convergence criterion.

\section{Calculation of the normalization constant, $A_{wall}$} \label{S:Awall}

Here we provide a brief description of the calculation of the normalization constant, $A_{wall}$ given in Eq.~\ref{eq:fwall3d}. From the form of Eq.~\ref{eq:fwall3d}, we can see that the wall field is uniform on the surface of a spherical geometry with uniform curvature. Accordingly, the normalization constant, $A_{wall}$ is chosen such that the strength of the wall field on the surface of a sphere obtained by the surface integral from Eq.~\ref{eq:fwall3d} is consistent with the wall field calculated from Eq.~\ref{eq:fwall2d} assuming that the polymer segment is located at the domain boundary (i.e, $d=0$ in Eq.~\ref{eq:fwall2d}). At the domain boundary, the strength of the wall field from Eq.~\ref{eq:fwall2d} is 
\begin{equation} \label{eq:fwalld0}
F_{ext}(\textbf{r}) = A_{0} \chi N [e^{2}-1]
\end{equation}

For a sphere of sufficiently large radius ($>>R_{g}$) and following Eq.~\ref{eq:fwall3d}, the strength of the wall field on the surface of the sphere can be approximately written as 
\begin{equation} \label{eq:fwalld03d}
F_{ext}(\textbf{r}) \sim 2 \pi A_{wall} \chi N \int _{0} ^{0.4R_{g}} rdr [e^{\frac{0.4 R_{g} - r}{0.2R_{g}}}] = 2\pi A_{wall} (0.4R_{g})^{2}\chi N \int _{0} ^{1} sds[e^{2(1-s)}]
\end{equation}
By equating the strengths of wall field in Equations ~\ref{eq:fwalld03d} and ~\ref{eq:fwalld0}, we get 
\begin{equation} \label{eq:awallval}
A_{wall} = A_{0}\frac{[e^{2}-1]}{2\pi (0.4R_{g})^{2}\int _{0} ^{1} sds[e^{2(1-s)}]} = A_{0}\frac{4[e^{2}-1]}{2\pi (0.4R_{g})^{2}[e^2-2]}
\end{equation}
It is to be noted that $A_{wall}$ is independent of the system size and hence, the strength of the wall field on the surface of a sphere of sufficiently large radius ($\sim$ a few times of $R_{g}$) will be similarly uniform and independent of the radius of the sphere.

\end{document}